\documentclass[11pt,preprint]{aastex}
\shorttitle{Resonant trans-neptunian objects}
\shortauthors{Gladman et al.}
\usepackage{natbib}
\begin{document}

\title{The Resonant Transneptunian Populations$^1$}

\author{
B. Gladman\altaffilmark{2},
S.M. Lawler\altaffilmark{2},
J-M. Petit\altaffilmark{3},
J. Kavelaars\altaffilmark{4},
R.L.~Jones\altaffilmark{5}, 
J.~Wm. Parker\altaffilmark{6}, 
C. Van Laerhoven\altaffilmark{2},
P. Nicholson\altaffilmark{7},
P. Rousselot\altaffilmark{3},
A. Bieryla\altaffilmark{6,8},
M.L.N. Ashby\altaffilmark{8}
}

\altaffiltext{1}{Based on observations obtained with MegaPrime/MegaCam, a joint
  project of CFHT and CEA/DAPNIA, at the Canada-France-Hawaii Telescope (CFHT)
  which is operated by the National Research Council (NRC) of Canada, the
  Institute National des Sciences de l'Universe of the Centre National de la
  Recherche Scientifique (CNRS) of France, and the University of Hawaii. This
  work is based in part on data products produced at the Canadian Astronomy
  Data Centre as part of the Canada-France-Hawaii Telescope Legacy Survey, a
  collaborative project of NRC and CNRS.}
\altaffiltext{2}{Dept.~of Physics \& Astronomy, 6224
  Agricultural Road, University of British Columbia, Vancouver, BC}
\altaffiltext{3}{Institut UTINAM, CNRS-UMR 6213, Observatoire de Besan\c{c}on,
  BP 1615, 25010 Besan\c{c}on Cedex, France}
\altaffiltext{4}{Herzberg Institute of Astrophysics, National Research
  Council of Canada, Victoria, BC V9E 2E7, Canada}
\altaffiltext{5}{Dept.~of Astronomy, Univ.~of Washington, Seattle WA, 98195}
\altaffiltext{6}{
Southwest Research
  Institute, 1050 Walnut Street, Suite 300, Boulder, CO 80302, USA}
\altaffiltext{7}{Cornell University, Space Sciences Building, Ithaca, New York
  14853, USA}
\altaffiltext{8}{Harvard-Smithsonian Center for Astrophysics, 60 Garden
  Street, Cambridge, MA 02138}

\begin{abstract}
The transneptunian objects (TNOs) trapped in mean-motion resonances with
Neptune were likely emplaced there during planet migration late in
the giant-planet formation process.
We perform detailed modelling of the resonant objects detected in
the Canada-France Ecliptic Plane Survey (CFEPS) in order to provide 
population estimates and, for some resonances, constrain 
the complex internal orbital element distribution.
Detection biases play a critical role because phase relationships
with Neptune make object discovery more likely at certain longitudes.
This paper discusses the 3:2, 5:2, 2:1, 3:1, 5:1, 4:3, 5:3, 7:3, 5:4,
and 7:4  mean-motion resonances, all of which had CFEPS detections, along
with our upper limit on 1:1 Neptune Trojans (which is consistent with their
small population estimated elsewhere).
For the plutinos (TNOs in the 3:2 resonance) we refine the orbital
element distribution given in Kavelaars {\it et al.} (2009) and show that steep
$H$-magnitude distributions 
($N(H)\propto 10^{\alpha H}$, with $\alpha=$0.8--0.9)
are favoured in the range $H_g$=8--9, and confirm that this resonance does 
not share the inclination distribution of the classical Kuiper Belt.
We give the first population estimate for the 5:2 resonance and find 
that, to within the uncertainties, the population is equal to that of 
the 3:2 ($\simeq$ 13,000 TNOs with $H_g<9.16$), 
whereas the 2:1 population is smaller by a factor of 3--4 compared 
to the other two resonances.
We also measure significant populations inhabiting the
4:3, 5:3, 7:3, 5:4, 7:4, 3:1, and 5:1 resonances, with 
$H_g<9.16$ ($D>$100~km) populations in the thousands.
We compare our intrinsic population and orbital-element distributions
with several published models of resonant-TNO production; the most
striking discrepancy is that resonances beyond the 2:1 are in reality
more heavily populated than in published models.

\end{abstract}

\section{Introduction}

The resonant transneptunian objects (TNOs) are a set of Edgeworth-Kuiper 
Belt objects whose orbital elements are such that the perturbations of Neptune 
causes relatively large-amplitude ($\sim$1\%) oscillations of the orbit 
on only 10$^4$-year time scales (much faster than secular oscillations in
the outer Solar System).
A necessary, but not sufficient condition for a object to be in a mean-motion
resonance is that its semimajor axis $a$ implies an orbital period $P$
which is a low-order integer ratio with Neptune with 
$P/P_N \simeq j/k$, where $j$ and $k$ are two small integers, 
in which case the object is said to be in the $j:k$ resonance\footnote{
The literature is mixed as to whether the periods or the mean-motions
should be the integer ratio, and thus some would call the external
resonance with twice Neptune's orbital period the 1:2 resonance.}
with Neptune, whose period is $P_N$ and semimajor axis $a_N$.
Kepler's 3rd law then provides the resonant semimajor axis 
$a = a_N (P/P_N)^{2/3}$.
Pluto was the first known resonant TNO; its presence in the 3:2 resonance
at $a\simeq$39.5~AU was discovered via direct numerical integration
\citep{CohenHubb1965}.
An important property of these resonances is that even resonant TNOs with
eccentricities $e$ so high that their perihelia $q$ satisfy
$q = a(1-e) < a_N$, and thus approach the Sun more closely than
Neptune, are `phase protected' by the resonance due to Neptune 
never being nearby 
when the TNO is at pericenter; 
in the case of Pluto, this phase protection means the planet actually
gets closer to Uranus than Neptune
\citep[although Pluto's orbit is
especially rich in resonant behaviors;][]{MilanietalPluto1989}.

In less than a year after the first moderately-sized TNOs began to
be discovered in the 1990's, other TNOs in the 3:2 resonance were 
recognized
\citep{Daviesetal2008}.
Termed `plutinos' \citep{JewittLuu1995}, these objects remain the
most numerous of the known resonant objects, with
\citet{Daviesetal2008}
reviewing the historical recognition of TNOs in other resonances.
The most recent compilations of accurately-measured resonant orbits
\citep{GladmanNomen2008,LykMuk2007} 
lists objects from the 1:1 (trojan) resonance all the way out 
to the 27:4 for 2004 PB$_{112}$ = I4212 as the current record
holder for largest resonant TNO semimajor axis, at $\simeq 108$ AU.
It is likely that even larger-$a$ resonant TNOs exist, but 
because the high-order resonances are thin in phase space, 
extremely-accurate orbits are required before the resonant behaviour 
can be confirmed.

A powerful idea is that the resonant TNOs were captured 
during an outward migration of Neptune in the distant past,
although there exist several contexts.
\citet{Malhot93} proposed Pluto's eccentricity had its origin
due to capture into the 3:2 as the resonance swept over the initial
heliocentric orbit of Pluto during Neptune's outward migration;
after capture, Neptune's continued migration forced up the captured
TNO's $e$ due to conservation of angular momentum.
\citet{HahnMalhot2005}
explored the sweep-up of resonant objects into a variety of resonances, 
showing how models match the observed-TNO distribution better if the 
resonances migrated into a primordial belt that has already been 
dynamically heated rather than the $e\sim i\sim 0$ case of a dynamically-cold
planetesimal disk, although achieving an inclination $i$ distribution as
hot as the observed objects was difficult.
\citet{Gomes2003} showed that abundant large-$i$ plutinos could be
produced if the plutinos were trapped out of a scattering disk 
already having interacted with Neptune, rather than from a 
pre-existing cold belt.
\citet{ChiangJordan2002}, \citet{ChiangDES2003}, and \cite{Ruth2005}
simulated resonant capture, looking at the population of resonances 
after the migration phase,
including studying how the relative populations of resonant `modes'
in the 2:1 resonance varied as a function of Neptune's migration 
distance and rate.
All these studies identified the problem that even though these models pump
eccentricities via the capture process, they still favoured migration 
into a dynamically pre-heated disk and even then the inclination 
distribution of the trapped resonant objects is not sufficiently high.
More recently, \citet{Levetal2008}
explored the idea that the {\it entire Kuiper Belt} was `planted' in
its current location as particles scattering off of Neptune during
the late stages of planet formation\footnote{The \citet{Levetal2008}
simulations are done in the context of the Nice model, which is usually
stated to be occurring ~600 Myr after Solar System formation.  However,
the Kuiper Belt implantation physics would work just as well if the
outward migration occurred in the few Myr following planetary formation.}
are dropped to lower eccentricity while temporarily trapped in 
mean-motion resonance, and are then decoupled into the current Kuiper belt; 
in this model the resonant particles are simply those that remained in 
the resonances at the end of migration. 
This model has several desirable properties, although the production of a 
Kuiper Belt with the correct inclination distribution is a challenge
\citep{L7paper}.
In our present manuscript, we will compare our measurements of how
various resonances are populated with some published models.

\subsection{Resonance dynamics}

\begin{figure}
\centering
\includegraphics[scale=0.35]{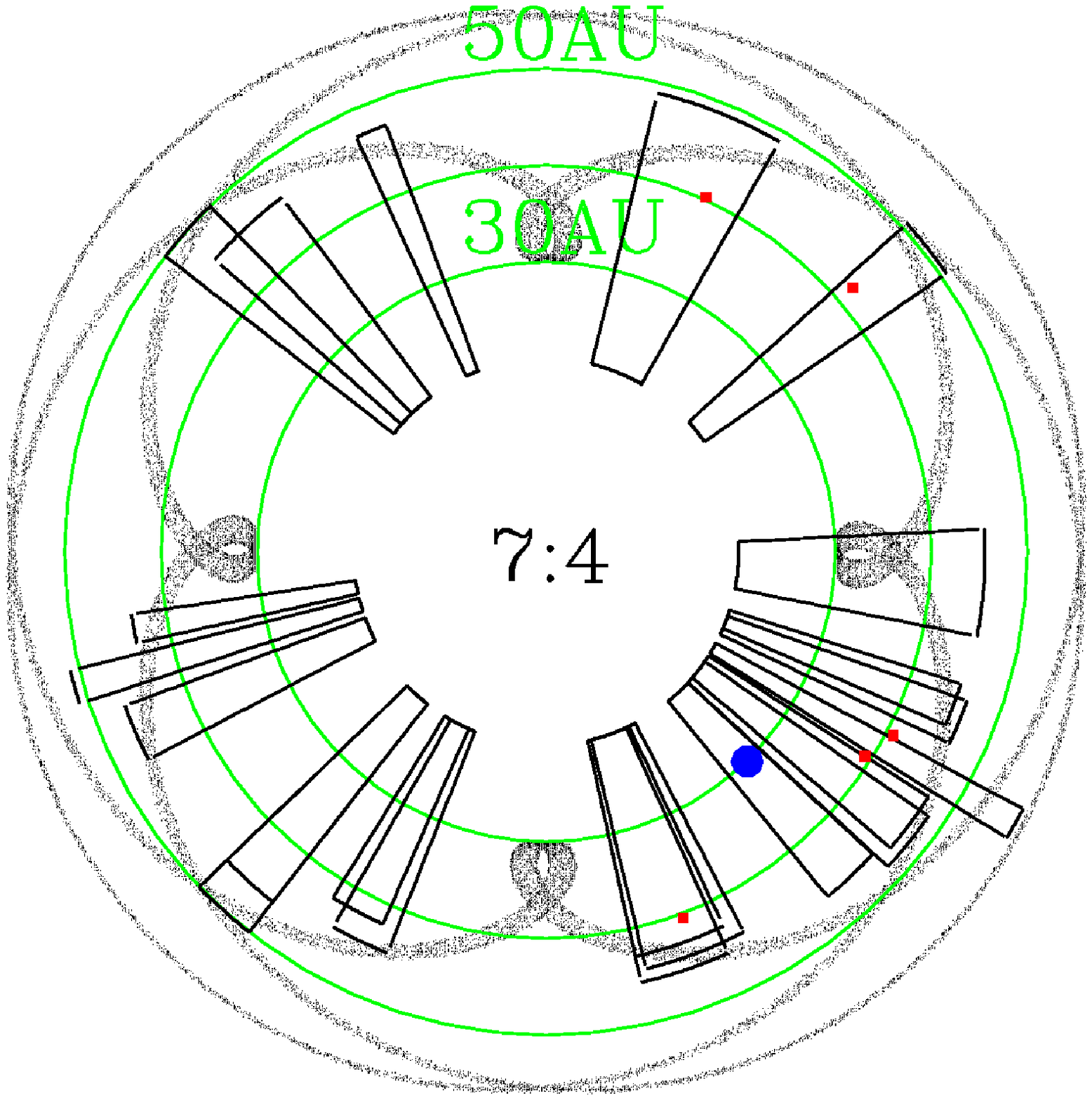}\includegraphics[scale=0.35]{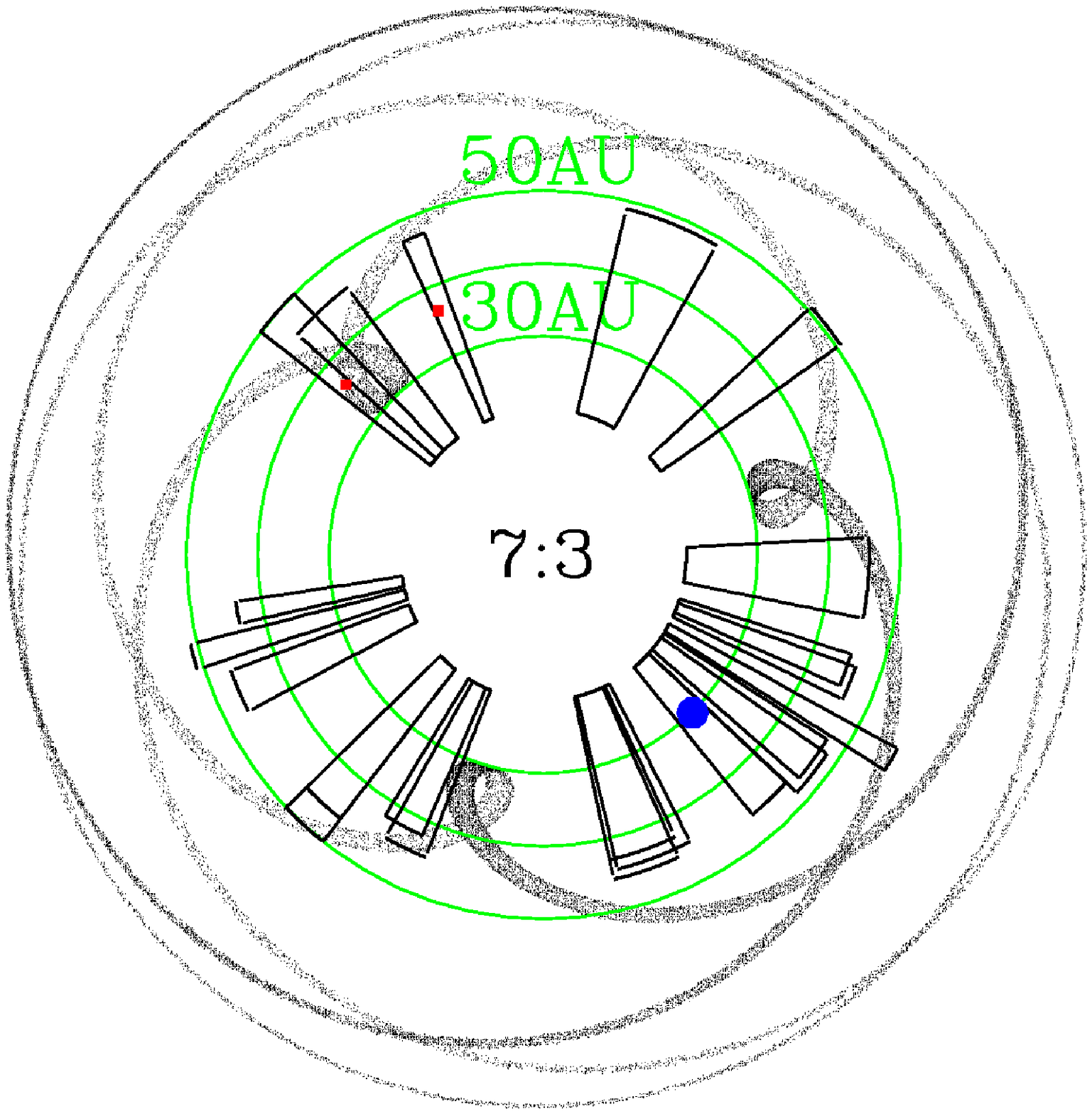}\includegraphics[scale=0.35]{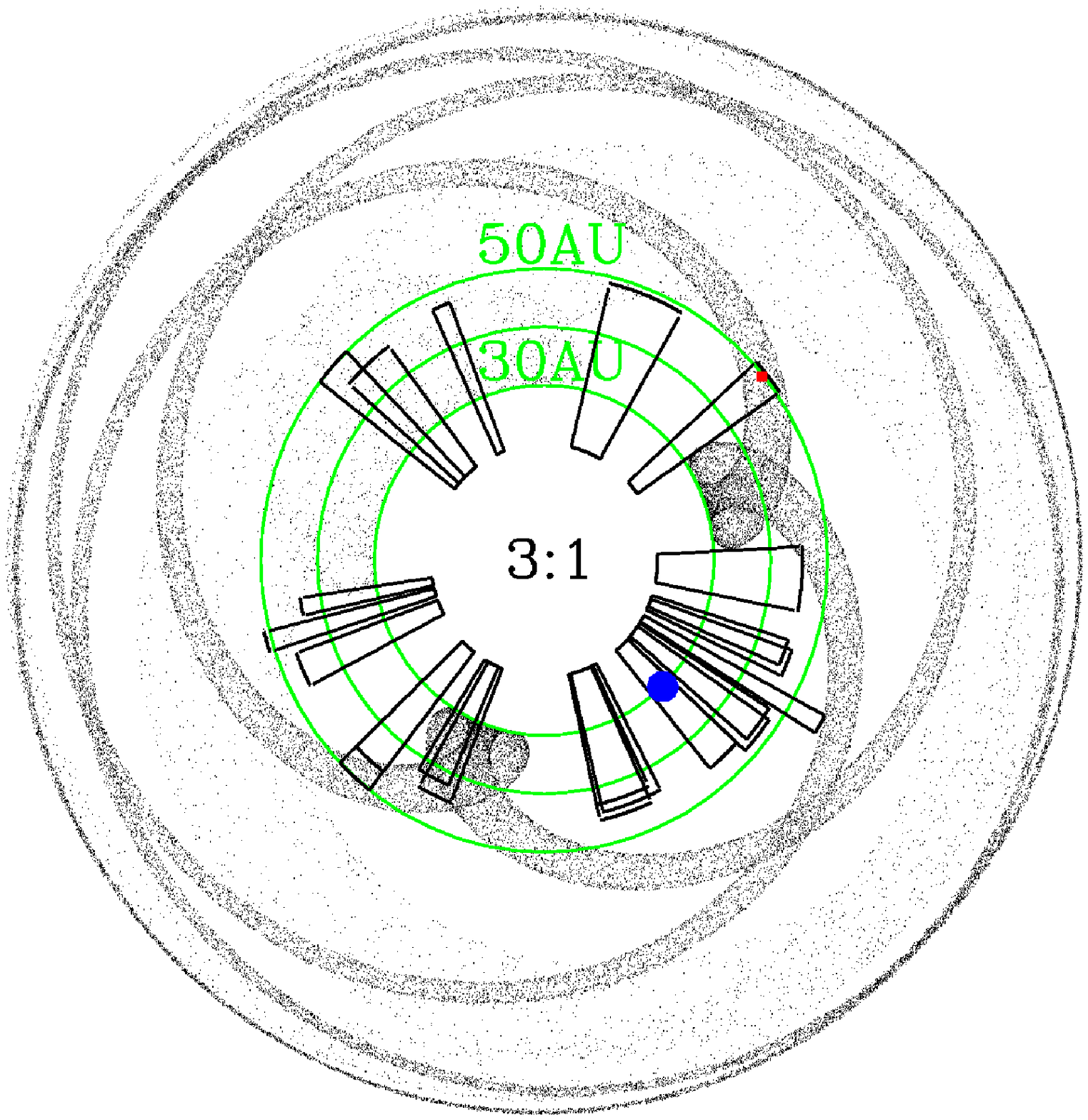}
\includegraphics[scale=0.35]{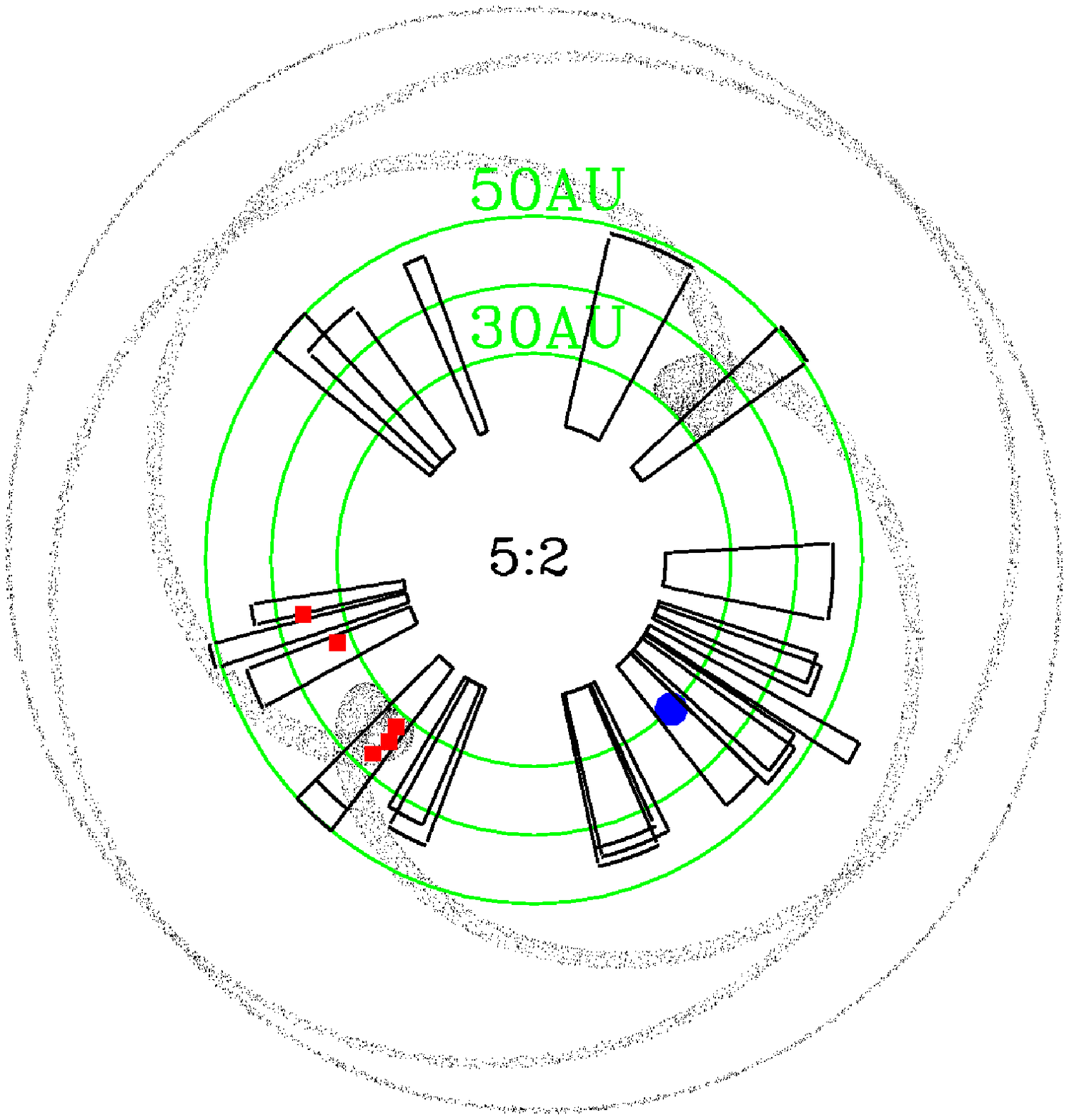}\includegraphics[scale=0.35]{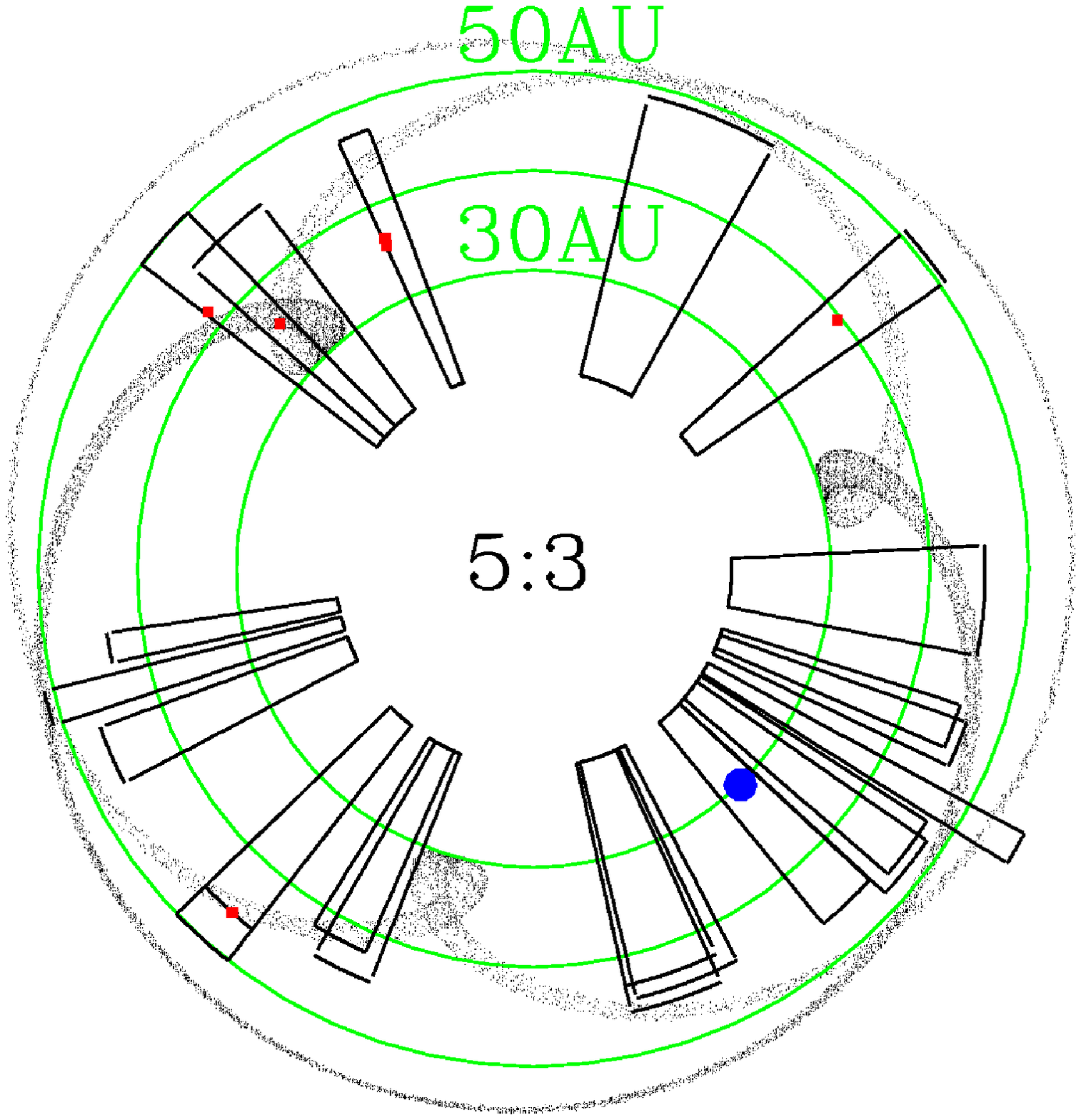}\includegraphics[scale=0.35]{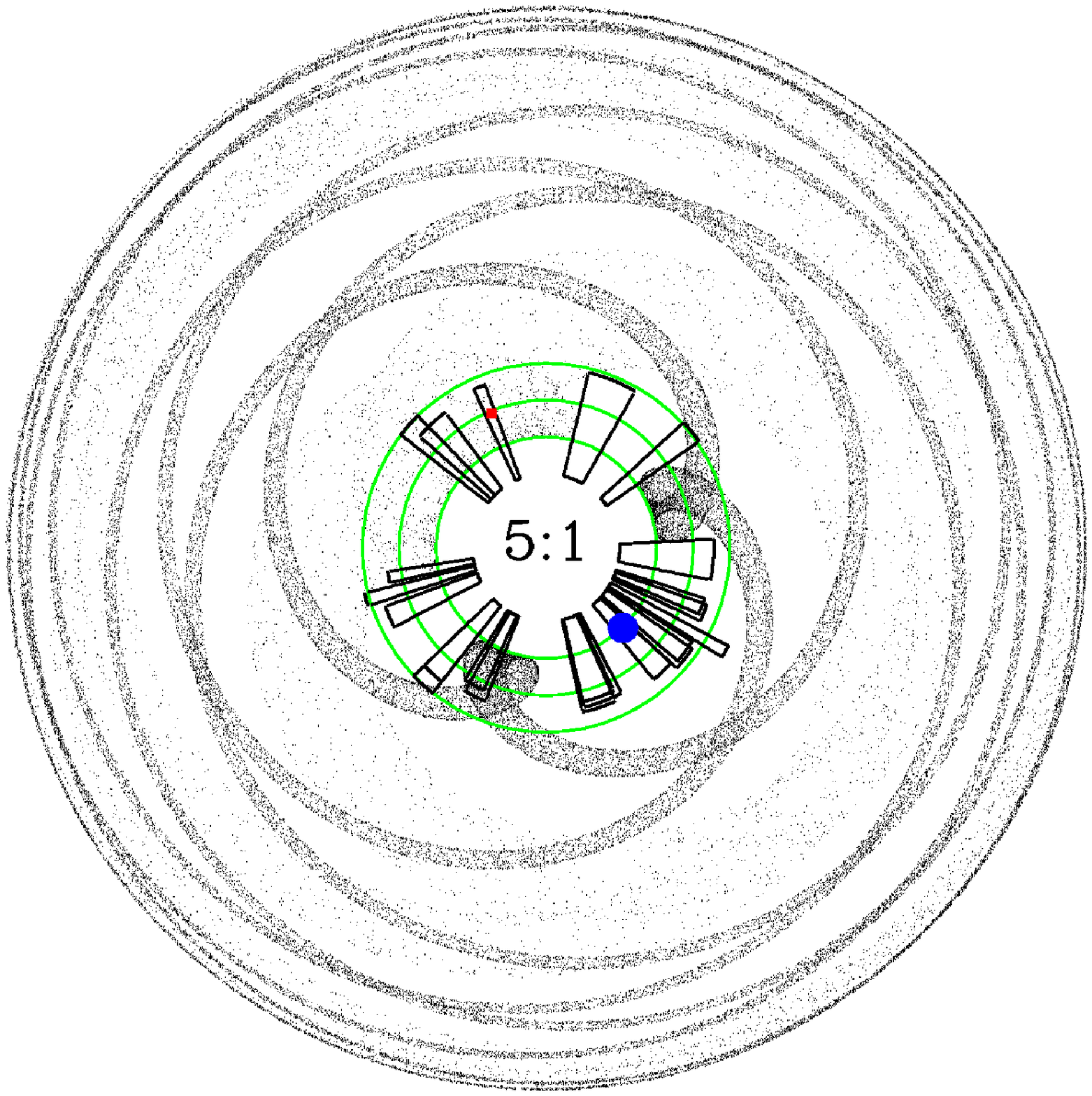}
\caption{
Toy models giving ecliptic projections (black dots) of TNOs 
with $i\simeq 0^\circ$, $q\simeq$30~AU, and libration amplitudes of 
$10^\circ$, to illustrate basic spatial TNO distribution induced by
a given resonance. 
These patterns stay fixed in the frame that co-rotates with Neptune, whose 
position is indicated by the large blue dot; green reference circles 
show heliocentric distances of $d$=30, 40, and 50~AU.  
Wedges show the ecliptic longitude range of the CFEPS blocks
(labelled in Fig.~\ref{fig:facedown32}), and 
red squares show the locations of the real CFEPS TNOs in that 
resonance.  
For the 3:1 and 5:1, 10\% of the model objects are in the symmetric 
libration island (with 50$^\circ$ libration amplitudes) and 45\% in each of the 
two asymmetric islands (with 10$^\circ$ amplitudes).
}
\label{fig:pretty}
\end{figure}

We provide only a brief tutorial on TNO resonant dynamics; further
introductory material can be found in 
\citet{MorbThomMoons},
\citet{Malhot1996},
\citet{ChiangJordan2002},
and \citet{GladKavCAP}.

Many TNOs are currently known to be 
in mean-motion resonances with
Neptune\footnote{No TNOs are yet securely known to inhabit mean-motion 
resonances with any other planets.}, 
meaning that the TNO's orbit is coupled to that of Neptune.
Neptune's mean longitude $\lambda_N$ (roughly its location around its orbit
as measured from the J2000 ecliptic reference axis) 
is related to the
TNO's longitude $\lambda$ (its current position) and the longitude
$\varpi$ of where the TNO's perihelion location is located.
Operationally,
inhabiting the $j:k$ resonance can be 
diagnosed by confirming (in a numerical integration) that the resonant
angle
\begin{equation}
\phi_{jk} = j \lambda - k \lambda_N - (j-k)\varpi
\label{eq:phijk}
\end{equation}
does not explore all values from 0 to 360$^\circ$.
The most common case (but not only possibility) for real resonant TNOs is 
that $\phi_{jk}$ oscillates (librates) around a mean
$<\!\!\phi_{jk}\!\!> = 180^\circ$ with some 
amplitude $L_{jk}$ (termed the libration amplitude).  
For example, a TNO in the 7:4 resonance with libration amplitude
$L_{74}=10^\circ$ means that $\phi_{74}$ oscillates (roughly sinusoidally)
between 170 and 190$^\circ$; such small amplitudes are rare in reality.
Because $\lambda=\varpi + {\cal M}$ where ${\cal M}$ is the TNO's mean anomaly, 
Eq.~(\ref{eq:phijk}) forces  that when the TNO is at perihelion (${\cal M}=0$),
\begin{equation}
\varpi - \lambda_N = \frac{1}{k} \; \phi_{jk} \; .
\label{eq:longdiff}
\end{equation}
In our example of the 7:4 resonance, this means that the TNO's pericenter
is `leading' ($\varpi - \lambda_N$) Neptune by (180/4)=45$^\circ$ for 
$<\!\!\phi_{74}\!\!\!>=180^\circ$; as $\phi_{74}$ oscillates by $\pm10^\circ$, the 
perihelion longitude oscillates by (10$^\circ$/4)=2.5$^\circ$ relative to the
45$^\circ$ offset (see the first panel of Fig.~\ref{fig:pretty}).
Because $\phi_{74}$ = 540$^\circ$, 900$^\circ$, and 1260$^\circ$ 
(adding multiples of 360$^\circ$ to 180$^\circ$) are all also valid,
this results in perihelion longitudes for libration center to be 
45$^\circ$, 135$^\circ$, 225$^\circ$, and 315$^\circ$ ahead of
Neptune for $L_{74}=0^\circ$ TNOs;
essentially one can add $2\pi m/k$ for any integer $m$ to the
right-hand side of (\ref{eq:longdiff}).
It is instructive to `trace the orbit' of a single  low-libration
amplitude TNO in the co-rotating panels of Fig.~\ref{fig:pretty}; any single
particle for the $j:k$ resonance explores all $k$ perihelion concentrations 
after making $k$ orbits around the Sun.
During that time Neptune will have made $j$ heliocentric
orbits.

The two rightmost panels of Fig.~\ref{fig:pretty} illustrate the
different generic case of the $n$:1 resonances, which can librate
in more that one state of perihelion locking relative to 
Neptune (these are usually called different `islands') despite
the fact that $k=1$ in Eq.~(\ref{eq:longdiff}).
Although these resonances still have `symmetric' libration of the
resonant argument $\phi_{n1}$ around an average value of 
$<\!\!\phi_{n1}\!\!>=180^\circ$, usually with very large
amplitude, they can also exhibit `asymmetric libration' 
around another $<\!\!\phi_{n1}\!\!>$ which depends on the value of the
orbital eccentricity
\citep{Beauge1994,Malhot1996}. 
Because these are $n:1$ resonances, the perihelion location of a
given such particle is confined to one of the two sky longitudes
(hence the term asymmetric); if the reader traces an asymmetric  3:1 orbit 
in Fig.~\ref{fig:pretty} they will see that it does {\it not} visit both
perihelion clusters.

The existence of confined pericenter locations for resonant TNOs has 
important implications for their observational study; surveys are
most sensitive to resonant TNOs that can be at perihelion in the
patch of sky being examined.
Because the number of TNOs increases rapidly as one goes to fainter
magnitudes (due to the size distribution being steep) and because most 
resonant TNOs occupy eccentric orbits ($e>0.1$ or much larger), the
number of detectable TNOs above the limit of a flux-limited survey
is a strong function of longitude relative to Neptune.
Essentially one becomes dominated by the hordes of smaller TNOs 
present near perihelion that become the majority of a detected sample.
Although this is a generic effect of eccentric populations
\citep{Jonesetal2006} 
it is more severe for the resonant populations than the main classical
belt due to the usually lower eccentricities of the latter;
we will illustrate this effect below with the plutino population.
The longitude bias in $\varpi$ shown in Fig.~\ref{fig:pretty}'s 
toy models are
more extreme than reality because the libration amplitude distribution
of the resonances is not concentrated towards zero.
This introduces yet another effect: during the oscillation of the resonant
argument more time passes with $\phi_{jk}$ near the extremes
\begin{equation}
\left| \phi_{jk} \right|_{extremal} \; = \;  <\phi_{jk}> \pm \; L_{jk}
\end{equation}
than at the libration center $ <\phi_{jk}> $ itself.
As an example, using the 5:3 resonance (Fig.~\ref{fig:pretty}), if
one were looking 85$^\circ$ ahead of Neptune, then the 5:3 resonators
(from the nearby `average' $<\phi_{53}>=60^\circ$ perihelion cluster)
with $L_{53} \simeq 75^\circ$ ($k=3$ times larger than the 25$^\circ$ 
longitude difference, {\it cf.,} Eq.~\ref{eq:longdiff})
will be favoured over other 5:3 resonators, 
if all other parameters are equal.

A similar detection bias is caused by orbital inclination $i$; as a TNO
rises and falls in ecliptic latitude it will spend less time at latitudes
near the ecliptic than at latitudes close to
$\pm i$.
This results in the true intrinsic TNO sky density of a given inclination 
peaking just below the latitude corresponding to the inclination, and
the ecliptic being the least likely place to find any given high-$i$
 TNO.

The real population in any given resonance is a superposition of all
eccentricities, libration centers, and libration amplitudes.
Conclusions about the distribution of any of these parameters cannot be
quantitative without detailed understanding of the longitude coverage
and depth of the surveys in which they  were found.

\section{Resonant CFEPS Objects}

The data acquisition of the CFEPS survey is described elsewhere 
\citep{Jonesetal2006,L3paper,L7paper}.
The survey coverage was divided into `blocks' of contiguous sky around
the ecliptic, labelled L3f through L7a, where the number indicates the
calendar year of the block's ``discovery'' observations (2003 through
2007) and the letter is the common MPC format designation of the 
two-week chunk to the calendar year (thus, discovery observations of
L5c were performed in the first half of February 2005).
Objects discovered in the block are given internal designations like
L5c11, indicating the eleventh TNO discovered in the L5c block.

This paper models the resonant CFEPS TNOs that are characterized 
detections\footnote{Characterized 
detections are those which have detection efficiencies $>$40\% in their CFEPS 
discovery block,
as defined in \citet{Jonesetal2006}.}
from the 3:2 (plutinos) and 5:2 resonances, three $n$:3 
resonances (the 4:3, 5:3, and 7:3), the 5:4 and 7:4 resonance, and 
three $n$:1 resonances (the 2:1, 3:1, and 5:1).
The orbital elements for the CFEPS TNOs in these resonances
are given in Tables~\ref{tab:elems} and \ref{tab:elems2}. 
In addition we give a 95\%-confidence upper limit on the 
Neptune Trojan population from our non-detection of such
an object.
Other resonances had zero or one CFEPS TNOs in them, and we elected
to not generate upper limits on their populations.

The discovery and tracking of these objects is discussed in
\citet{L7paper}.
Important for our purposes here is:
(1) a wide range of ecliptic longitudes were surveyed with CFEPS,
    which means CFEPS was sensitive to objects with a large variety 
    of libration amplitudes,
(2) patches of sky away from the perihelion longitudes of the resonances
    were quantitatively characterized; the {\it non-detection} of
    resonant objects at those longitudes provides powerful constraints
    on the large-amplitude resonators, and
(3) an extremely high fraction of the discoveries were tracked, preventing
    loss of unusual objects, as described in \citet{Jones2010}.
    As an example of this, CFEPS re-discovered 
    TNOs actually inhabiting the rare 5:4 and 7:3 mean-motion resonances
    (L3y11, L3y07 and L5c19PD; see Table~\ref{tab:elems2} caption) 
    at on-sky positions $\sim1^\circ$ from the ephemeris that 
    had been assigned based on an incorrect orbit computed from the 
    short-arc discovery prior to 2003 ({\it i.e.,} the TNOs had been 
    lost before their resonant nature was recognized).

Tables~\ref{tab:res} and \ref{tab:res2} list the current barycentric 
$a,e,i$ J2000 osculating elements of each object and a determination of the 
resonant libration amplitude,
which comes from the range of possible orbits as diagnosed in the method
of \citet{GladmanNomen2008}.
The resonance amplitudes listed should be interpreted as a range
which encompasses nearly all ($>$99\%) of the possible true values
of the TNO's libration amplitude.
Because the CFEPS tracking strategy regularly provided
off-opposition observations during the 3-opposition orbits, the libration
amplitudes for the CFEPS sample are more precise than for the majority of the
MPC sample given in \citet{GladmanNomen2008} and 
\citet{LykMuk2007} 
because many of the objects in the MPC database have much sparser astrometric
coverage. 

In addition to mean-motion libration amplitudes, the Kozai resonance 
(see Appendix) is 
observed to function for two CFEPS plutinos (L4h07 and L4k01); 
Table~\ref{tab:elems} gives the amplitude and mean value of the argument 
of pericenter $\omega$ (which is effectively the resonant angle).
For objects in $n$:1 resonances where there are symmetric and asymmetric
libration islands, Table~\ref{tab:elems2} identifies the mode and estimates
of the libration center position and libration amplitude.

There are a few high-order resonances with CFEPS detections which we
do not model here due to the fact that the resonant occupation is not
yet secure.
These include the 15:8, 17:9, and 12:5 mean-motion resonances,
and are listed as insecure resonators in  
\citet{L7paper}.


\begin{deluxetable}{lr|llrrrccc} 
\tabletypesize{\scriptsize}
\tablecaption{CFEPS 3:2 (plutinos) and 5:2 Resonators
\label{tab:res}}
\tablehead{
\multicolumn{2}{c|}{DESIGNATIONS} 
& \colhead{a}    & \colhead{e} &   \colhead{i}    & \colhead{d} 
& \colhead{res}  & \colhead{amp}  & \colhead{mag} & \colhead{Comment} 
\\
\colhead{CFEPS} & \colhead{MPC} & \colhead{(AU)} & & \colhead{($^\circ$)} & \colhead{(AU)} 
& &  \colhead{($^\circ$)} & \colhead{($g$)} & .
}
\startdata
 L4k11 & 2004 KC19 &  39.258 & 0.23605 &  5.637 & 30.2 &  3:2 &  79$\pm$23 & 23.3 & \\ 
 L4h15 & 2004 HB79 &  39.260 & 0.22862 &  2.661 & 32.0 &  3:2 &  82$\pm$13 & 24.0 & \\ 
 L5c11 & 2005 CD81 &  39.262 & 0.15158 & 21.344 & 45.2 &  3:2 &  98$\pm$ 7 & 23.7 &\\ 
 L4h06 & 2004 HY78 &  39.302 & 0.19571 & 12.584 & 31.8 &  3:2 &  74$\pm$ 8 & 23.8 & \\ 
 L4v18 & 2004 VY130&  39.342 & 0.27616 & 10.203 & 28.5 &  3:2 &  38$\pm$19 & 23.3 & \\ 
 L4m02 & 2004 MS8  &  39.344 & 0.29677 & 12.249 & 27.8 &  3:2 & 125$\pm$ 2 & 23.4 & \\ 
 L3s02 & 2003 SO317& 39.346 & 0.2750  &  6.563 & 32.3  &  3:2 & 100$\pm$20 & 23.8 & \\
 L4h09PD&  47932   & 39.352 & 0.28120 & 10.815 & 28.5 &  3:2 &  54$\pm$15 & 21.3 & \\ 
 L3h19 & 2003 HF57 & 39.36  & 0.194   &  1.423 & 32.4  &  3:2 & 60$\pm$20 & 24.2 & \\
 L3w07 & 2003 TH58 & 39.36  & 0.0911  & 27.935 & 35.8  &  3:2  & 100$\pm$10& 23.0 & \\
 L4h07 & 2004 HA79 &  39.378 & 0.24697 & 22.700 & 38.4 &  3:2 &  46$\pm$11 & 23.7 & Kozai 270$\pm30^\circ$ \\ 
 L3h11 & 2003 HA57 & 39.399 & 0.1710 & 27.626 & 32.7  &  3:2  &  70$\pm$ 5 & 23.4 & \\
 L3w01 & 2005 TV189 & 39.41  & 0.1884  & 34.390 & 32.0 & 3:2  &  60$\pm$20 & 22.9 & \\
 L4j11 & 2004 HX78 &  39.420 & 0.15270 & 16.272 & 33.6 &  3:2 &  28$\pm$ 5 & 23.6 & \\ 
 L4v09 & 2004 VX130 &  39.430 & 0.20696 & 5.745 & 34.8 &  3:2 &  50$\pm$32 & 23.5 & \\ 
 L3h14 & 2003 HD57  & 39.44  & 0.179   &  5.621 & 32.9 &  3:2 &  60$\pm$30 & 23.3 & \\
 L3s05 & 2003 SR317 & 39.44  & 0.1667  &  8.348 & 35.5 &  3:2 &  90$\pm$ 5 & 23.7 & \\
 L4v13 & 2004 VV130 & 39.454 & 0.18827 & 23.924 & 32.8 &  3:2 &  49$\pm$12 & 22.7 & \\ 
 L4k01 & 2004 KB19 &  39.484 & 0.21859 & 17.156 & 39.5 &  3:2 &  57$\pm$31 & 24.0 & Kozai 270$\pm50^\circ$ \\ 
 L3h01 & 2004 FW164 & 39.492 & 0.1575  &  9.114 & 33.3 &  3:2 &  80$\pm$20 & 23.8 & \\
 L5i06PD & 2001 KQ77& 39.505 & 0.15619 & 15.617 & 36.2 &  3:2 &  72$\pm$ 8 & 23.1 & \\ 
 L4h10PD & 1995 HM5 & 39.521 & 0.25197 &  4.814 & 31.1 &  3:2 &  77$\pm$20 & 23.8 & \\ 
 L4v12 & 2004 VZ130 & 39.551 & 0.28159 & 11.581 & 29.2 &  3:2 &  88$\pm$10 & 24.0 & \\ 
 L4h08 & 2004 HZ78 & 39.580 & 0.15095 & 13.310 & 34.8 &  3:2 & 115$\pm$15 & 23.0 & \\ \hline
 L4j08 & 2004 HO79 &  55.206 & 0.41166 &  5.624 & 37.3 &  5:2 &  84$\pm$20 & 23.5 & \\ 
 L3f04PD & 60621   & 55.29  & 0.4020  &  5.869 & 36.0  &  5:2  & 80$\pm$30 & 22.7 & \\ 
 L4j06PD & 2002 GP32 &  55.387 & 0.42195 &  1.559 & 32.1 & 5:2 & 65$\pm$ 2 & 22.1 & \\ 
 L4k14 & 2004 KZ18 &  55.419 & 0.38191 & 22.645 & 34.4 &  5:2 &  44$\pm$10 & 24.1 & \\ 
 L4h02PD & 2004 EG96 &  55.550 & 0.42291 & 16.213 & 32.2 & 5:2 & 91$\pm$17 & 23.5 & \\  \hline
\enddata
\tablecomments{Characterized CFEPS resonators, with MPC (where available) 
designations.
A `PD' suffix indicates that the CFEPS team realized immediately
that this was a previously-discovered TNO, but which could now be used
in our flux-calibrated analysis.
All digits in the best-fit barycentric orbital $a/e/i$ are significant.
$g$-band magnitudes are rounded to 0.1 mags, with exact values and errors
given in Table 7 of \citep{L7paper}.
Heliocentric distances $d$ and $H_g$ magnitudes are given at the
first date of CFEPS detection.
Libration amplitude is the best fit orbit's value along with the range 
covering $>$99\% of possible values given orbital uncertainties.
For Kozai librators the libration center of $\omega$ and amplitude 
$A_\omega$ are given.
}
\label{tab:elems}
\end{deluxetable}

\begin{deluxetable}{lr|llrrrccc} 
\tabletypesize{\scriptsize}
\tablecaption{  CFEPS TNOs in Resonances Other than 3:2 and 5:2
\label{tab:res2}
}
\tablehead{
\multicolumn{2}{c|}{DESIGNATIONS} 
& \colhead{a}    & \colhead{e} &   \colhead{i}    & \colhead{d} 
& \colhead{res}  & \colhead{amp}  & \colhead{mag} & \colhead{Comment} 
\\
\colhead{CFEPS} & \colhead{MPC} & \colhead{(AU)} & & \colhead{($^\circ$)} & \colhead{(AU)} 
& &  \colhead{($^\circ$)} & \colhead{($g$)} & .
}
\startdata
 L3y11 & 131697    & 34.925 & 0.0736  &  2.856 & 34.0  &  5:4 &  75$\pm$ 5 & 23.8 & MPC$_W$   \\ \hline
 L4h14 & 2004 HM79 &  36.441 & 0.07943 &  1.172 & 38.0 &  4:3 &  63$\pm$ 1 &  23.7 & \\ 
 L3s06 & 143685    & 36.456 & 0.2360  &  5.905 & 28.2  &  4:3 &  60$\pm$20 & 22.8 & \\
 L5c23 & 2005 CF81 &  36.473 & 0.06353 &  0.405 & 34.4 &  4:3 &  47$\pm$15 & 24.2 & \\ 
 L7a10 & 2005 GH228&  36.663 & 0.18814 & 17.151 & 30.6 &  4:3 & $\sim$120 & 23.6 & Insecure\\  \hline
 L5c08 & 2006 CJ69 &  42.183 & 0.22866 & 17.916 & 35.5 &  5:3 &  70$\pm$40 & 23.6 & \\ 
 L3y06 & 2003 YW179& 42.193 & 0.1537  &  2.384 & 35.7  &  5:3 & 100$\pm$20 & 23.7 &  \\
 L5c13PD&2003 CX131& 42.240 & 0.23387 &  9.757 & 41.8 &  5:3 &  72$\pm$ 3  & 23.8 & \\ 
 L4v05 & 2004 VE131&  42.297 & 0.25889 &  5.198 & 39.6 &  5:3 &  81$\pm$31 & 24.1 & \\ 
 L3y12PD & 126154  & 42.332 & 0.14043 & 11.078 & 36.4  &  5:3  & 100$\pm$10& 21.7 & \\
 L4k10 & 2004 KK19 &  42.410 & 0.14391 &  4.485 & 46.0 &  5:3 & $\sim$125  & 24.4 & Insecure\\ \hline
 L3q08PD & 135742  & 43.63  & 0.125   &  5.450 & 40.7  &  7:4  & $\sim$60  & 23.7 & \\ 
 L4n03 & 2004 OQ15 &  43.646 & 0.12472 &  9.727 & 40.5 &  7:4 & $\sim$60   & 23.7 & \\ 
 L3w03 & 2003 YJ179& 43.66  & 0.0794  &  1.446 & 40.3  &  7:4  & $\sim$130 & 23.8 & \\
 L4v10 & 2004 VF131&  43.672 & 0.21492 &  0.816 & 42.0 &  7:4 & $\sim$90   & 23.9 & \\ 
 K02O03& 2000 OP67 &  43.72  & 0.191   &  0.751 & 39.3 &  7:4 & $\sim$70   & 24.3 &  \\ \hline
 L4h18 & 2004 HP79 &  47.567 & 0.18250 &  2.253 & 39.5 &  2:1 & $\sim$50   & 23.3 & asym. $\sim$258 \\ 
 L4k16 & 2004 KL19 &  47.660 & 0.32262 &  5.732 & 32.3 &  2:1 &  20$\pm$ 7 & 24.0 & asym. 288$\pm$1 \\ 
 L4k20 & 2004 KM19 &  47.720 & 0.29180 &  1.686 & 33.8 &  2:1 &  12$\pm$ 4 & 23.8 & asym. 287$\pm$1 \\ 
 K02O12&2002 PU170 &  47.75  & 0.2213  &  1.918 & 47.2 &  2:1 & 154$\pm$ 4 & 24.3 & symm. \\ 
 L4v06 & 2004 VK78 &  47.764 & 0.33029 &  1.467 & 32.5 &  2:1 &  23$\pm$ 5 & 23.7 & asym.  73$\pm$1\\ \hline
 L3y07   & 131696  & 52.92  & 0.3221  &  0.518 & 36.6  &  7:3 & 100$\pm$20 & 23.4 & MPC$_W$   \\
 L5c19PD&2002 CZ248&  53.039 & 0.38913 &  5.466 & 36.2 &  7:3 &  84$\pm$20 & 23.8 & MPC$_W$ \\ \hline
  L4v08 & 2004VD130& 62.194 & 0.42806&  8.024 & 49.7 &  3:1 & $\sim$160 & 24.0 & symmetric? \\ \hline
  L3y02 &2003 YQ179&  88.38 & 0.5785 & 20.873 & 39.3 &  5:1 & $\sim$160 & 23.4 & Insecure, symmetric\\ 
\enddata
\tablecomments{Characterized 
CFEPS and MPC (where available) designations are given;
objects beginning with 'K' are from the CFEPS presurvey \citep{Jonesetal2006}.
All digits in the best-fit barycentric J2000 orbital $a/e/i$ are significant.
Heliocentric distances $d$ at detection are rounded to 0.1 AU.
$g$-band magnitudes are rounded to 0.1 mags, with exact values and errors
given in Table 7 of \citep{L7paper} or \citep{Jonesetal2006} (the latter 
assuming $g-R$=0.8).
Libration amplitudes are the range covering $>$99\% of possible true orbits.
For $n:1$ resonances the libration island and mean-resonant argument are given.
'Insecure' indicates that this resonance occupation is not secure according
to the \citet{GladmanNomen2008} criterion.
`MPC$_W$' indicates the TNO was in MPC database with the wrong orbit; CFEPS
re-found the objects (usually $>1^\circ$ from predicted location) and
CFEPS discovery and tracking observations improved the orbit to the
listed values.
}
\label{tab:elems2}
\end{deluxetable}

\section{CFEPS survey simulation of a resonant population}

We model the orbital distribution in each resonance with several 
goals.
The orbital element distribution inside each resonance is represented
either by a parametric model or, in the case of the $n$:1 resonances,
a prescription based on the known dynamics of the resonance.
In the case of a parametric model the functional forms chosen are ones
which post-facto provide a reasonable match between the simulated 
and real CFEPS detections.
In order to converge to our best models, candidate orbital distributions 
were tested as described in \citet{L3paper} and the best-matching
models were determined; briefly, models for which one of the $e, i, d, m_g,$ 
or $L$ cumulative distributions have an Anderson-Darling statistic 
which occurs by random $<$5\% of the time are rejected.
For example, we find that for most of the resonances the intrinsic
eccentricity distribution can be satisfactorily represented by a
probability distribution in the form of a gaussian with center
at eccentricity $e_c$ and half-width $e_w$ (rejecting negative
eccentricities).
These parametric representations are entirely empirical, due to the
fact that there is no physical model that provides a parametric
form.
However, because these functional forms provide rather satisfactory
matches to the CFEPS detections, theoretical models of resonant TNO production
will have to provide orbital parameter distributions that give roughly the 
same distribution as our intrinsic model, rather than values in the 
biased MPC sample.
For example, we find the plutino eccentricity distribution is strongly
peaked near $e_c$=0.18 with narrow width; this intrinsic $e_c=0.18$ peak is
below the median plutino $e$ of 0.22 in the Minor Planet Center.
Similarly, we find the median intrinsic plutino inclination to be 
$\approx16^\circ$, whereas detected samples from ecliptic surveys (biased
towards low-inclination detections) have median inclinations $\sim12^\circ$
both for CFEPS and the Deep Ecliptic Survey 
\citep[abbreviated DES hereafter,][]{Gulbis2010}. 

Our second goal is to produce debiased population estimates for 
each resonance, in order to compare the resonances to each other
and to other Kuiper Belt components.
For many resonances we lack sufficient detected numbers to explore
the internal orbital distribution in detail, but can nevertheless 
provide calibrated absolute population estimates which should be accurate 
to a factor of a few, based on analytic expectations of the resonance's 
internal structure,

The CFEPS Survey Simulator begins with synthetic objects having a range
of $H_g$ magnitudes and with orbital elements that place them within a 
given resonance, correctly time weighted for their occupation of 
different regions of phase space.  
Due to differing structure, orbital elements for each resonance's
simulated objects are chosen differently;  
the procedure for each of the three groups of resonances are described 
in the Appendix.
As each synthetic object is created, 
the CFEPS pointings, magnitude limits, and tracking efficiencies are
applied, to decide whether or not the object is detected.  
New synthetic objects are created and checked for detectability 
until a user-defined number of synthetic detections are
acquired.  
If this desired number is equal to the number of CFEPS detections,
the simulation provides an estimate of the intrinsic population of 
the resonance.
If instead a cosmogonic model is available, a large number of 
synthetic detections may be requested, in order to build a well-sampled
distribution of the orbital elements that the cosmogonic model
predicts CFEPS should detect.
The orbital elements (eccentricity, inclination, discovery distance, 
apparent magnitude, and libration amplitude)
of these synthetic detections are then compared
statistically to the real detections to determine whether or not our 
distribution of synthetic detections from that model is rejectable.


\section{The Plutinos (3:2 resonators)}

\begin{figure}
\centering
\includegraphics[scale=0.65]{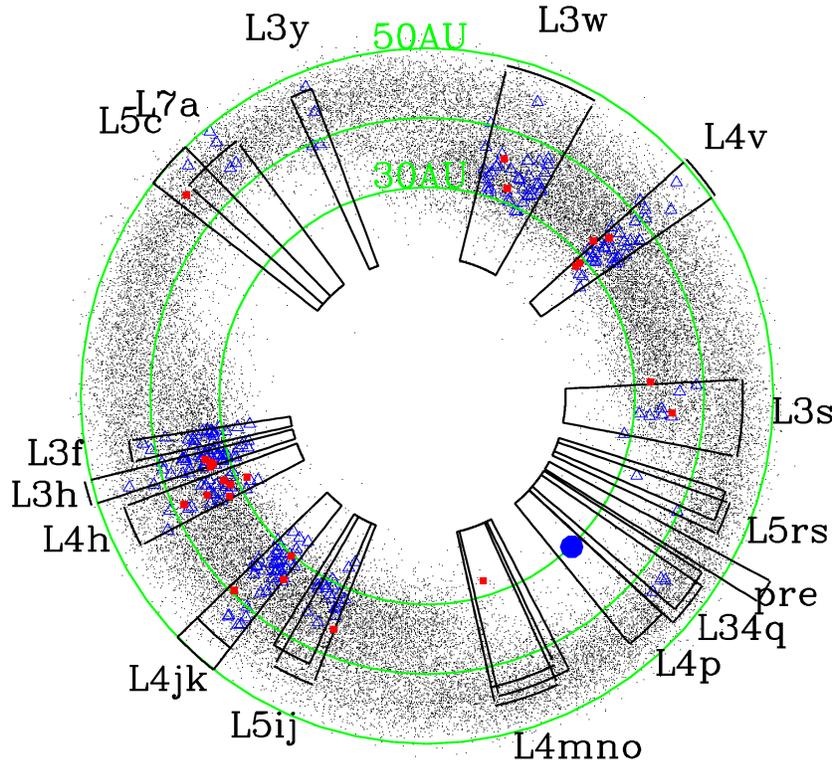}
\caption{Ecliptic projection of the plutinos.  
The filled red squares are the 24 real detected plutinos, 
open blue triangles are 240 simulated detections, 
and tiny black dots show our model's intrinsic plutino 
distribution. 
Neptune's position is shown by 
the large blue dot.
The CFEPS `blocks' are shown as wedges covering the correct ecliptic
longitude range, where the inner edges at $\sim$20~AU are set by the 
detection pipeline's  rate cut and the outer extent is at the distance
where a $H_g$=7.5 TNO would cease to be visible (larger TNOs are visible 
further away of course).
The syntax L4jk means the L4j and L4k blocks are overlapping.
The two ecliptic intersections with the galactic plane are roughly
straight up and down in this diagram.
}
\label{fig:facedown32}
\end{figure}

The plutinos (3:2 mean-motion librators) are by far the largest
sample in the flux-biased catalogues.
This preponderance is partly due to the low semimajor axis,
keeping heliocentric distances $d$ low, but detection of objects in $n$:2 
resonances is also favoured over many other resonances because their perihelion
sky densities are currently (due to Neptune's position over the last
two decades) larger at the high galactic latitudes that Kuiper
Belt surveys have tended to favor.
This well-known effect is illustrated in
Fig.~\ref{fig:facedown32}
which shows the CFEPS survey block locations along with the CFEPS plutinos
discovered (and tracked to obtain orbits with $\delta a/a < 10^{-4}$).

We began by improving the nominal CFEPS L3 plutino model, using
the tripled sample size of 24 CFEPS detections (8 of which were
part of the L3 plutino sample).
To our surprise, the orbital distribution settled on by \citet{L3paper}
from only eight characterized L3 plutinos remains a non-rejectable
model despite tripling the sample, showcasing the ability of 
well-characterized surveys to constrain orbital distributions.
Although we cannot reject the L3 plutino model, the 24 CFEPS plutinos
now allow us to improve the details of the plutino model to explore
other aspects of the resonance that were not accessible with a sample
size of eight.

\subsection{The plutino inclination distribution}

\begin{figure}
\centering
\includegraphics[scale=0.6]{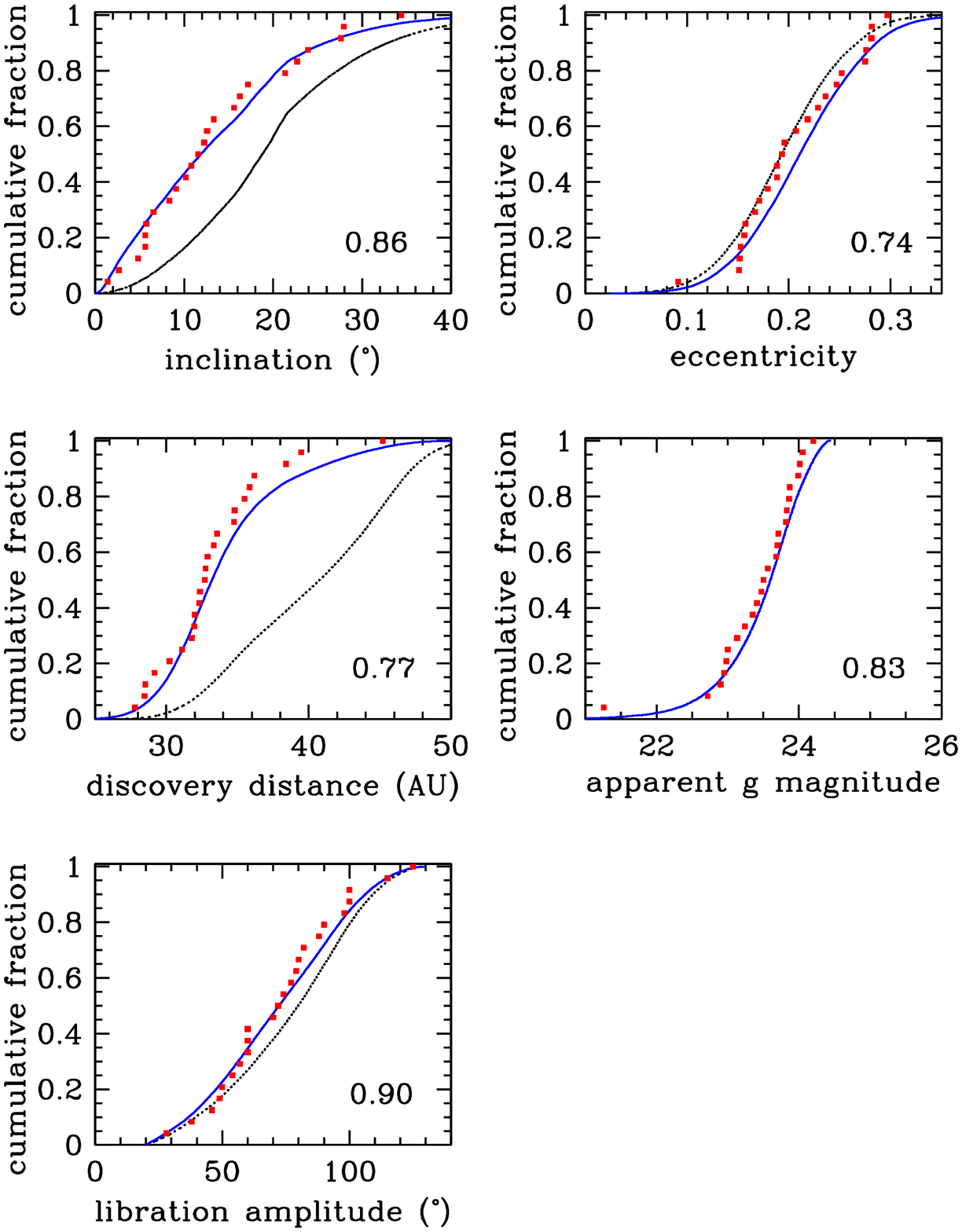}
\caption{Cumulative distributions of the 5 variables on which we perform
statistical analysis, for the plutinos.  
Red squares show the distribution of the 24 detected CFEPS 
characterized plutinos.
The dashed black line shows the distribution of the intrinsic plutino
population from our favored L7 model,
and the thicker blue line shows resulting distribution of that model's 
simulated detections.
The number in each panel is the bootstrapped Anderson-Darling
statistic, indicating the percentage of randomly-drawn samples from the
simulated detection distribution that had worse Anderson-Darling values 
than the real detections.  
We reject the model if any
parameter has a bootstrapped value $<$0.05 (meaning 
only 5\% of randomly-drawn samples have a worse Anderson-Darling 
statistic than the real detections).
}
\label{fig:cumuplots32}
\end{figure}

We find that an orbital-inclination probability distribution of the form 
\begin{equation}
P(i) \propto \sin i \exp \left( \frac{-i^2}{2\sigma_{32}^2} \right)
\end{equation}
provides an acceptable representation of the intrinsic plutino inclination,
with $\sigma_{32}=16^\circ$ giving the best match
(first panel of Fig.~\ref{fig:cumuplots32}a).
The CFEPS 95\% confidence intervals for this functional form range
from $12^\circ$ to $24^\circ$.
The lower end of this range overlaps with the estimates of 
$\sigma_{32} = 10^{+3}_{-2}$ degrees \citep{Brown2001}
and $\sigma_{32} = 11\pm2^\circ$ \citep{Gulbis2010}, which use a 
heavily-overlapping sample.
As was the case in \citet{L3paper}, the CFEPS survey continues to
favour a significantly-hotter inclination distribution for the 
plutinos, with 4 of our 24 plutinos having $i>23^\circ$, whereas
none of the 51 DES plutinos have $i>23^\circ$.
We do not believe this is a sample-size problem, but rather an
issue of preferential loss of the large-inclination detections in
surveys that did not systematically acquire tracking observations
2--4 months after discovery in the initial opposition; 
\citet{Jones2010} illustrates how this bias enters Kuiper Belt
surveys, regardless of the orbit fitting method used for the 
short-arc orbits.
Since plutinos are often discovered at nearby 30--35~AU distances,
their faster rate of motion makes accurate determination of their
orbits more critical than classical objects, and they are easier to
lose at the next opposition.

Our plutino inclination distribution is quite similar to the
inclination distribution of the `hot' component of the classical
Kuiper Belt, making it plausible that the plutinos and the 
hot classical belt are both captured populations whose inclination
distribution neither affected their capture probability, nor was
$i$ critical for post-capture erosion over the Solar System's
age.
We have shown that trying to use the same bimodal inclination for the 
plutinos as for the classical belt yields rejection at far more 
than 99\% confidence.

We also explored a functional form of
$
P_2(i) \propto \sin^2 i \; \exp \left( \frac{-i^2}{2\sigma_s{^2}} \right)
$
for the plutinos, which is roughly a Maxwellian distribution for the
velocity component perpendicular to the plane. 
This functional form also gives perfectly acceptable matches to
the CFEPS detections, with a best match at $\sigma_s=11^\circ$
and an acceptable range (95\% confidence) from 
$\sigma_s$=8.5--13.5$^\circ$.
However, because this parameterization did not give a significantly better
match, nor did it change the total population estimates by more than 
their uncertainties, for ease of comparisons with the literature
we have elected to retain the $\sin (i)$, rather than $\sin^2 (i)$, 
formulation. 

We checked our plutino sample's colors,
tabulated in \citet{L7paper}, for a correlation with inclination
or `size' via $H_g$ \citep{Almeida2009,RuthHilke2011}, but
find no significant correlation.
We postulate that the size versus inclination correlation is an
artifact of the survey depths that found them (with shallow
wide-area surveys finding essentially all $H<5$ plutinos far
from the ecliptic, whereas most fainter plutinos have been
found in ecliptic surveys which don't cover enough area to
find the few $H<5$ TNOs near the ecliptic).
Our plutino sample does not quite go deep enough (past $H\sim 8.5$)
to have enough discrimination to see if the smaller plutinos
become bluer; our colors are all uniformly blue.

\subsection{The plutino Kozai subcomponent}

Two (8\%) of our 24 CFEPS plutinos (L4h07 and L4k01) are also in
the Kozai resonance; their Kozai classifications are secure using the
\citet{GladmanNomen2008} nomenclature.
The argument of pericenter for a Kozai plutino librates around 
$\omega$=90 or 270$^\circ$ due to the fact that the resonance affects 
the angular precession rate, so although \citet{ThomMorbKoz} show that 
the Kozai effect in the non-resonant Kuiper Belt appears only at large
$e$ and $i$, inside the 3:2 resonance the Kozai resonance can 
appear for even moderate-inclination plutinos \citep{MorbThomMoons}.
The libration amplitude $A_\omega$ depends on the initial $e$, $i$,
and $\omega$.
The plutinos
L4h07 and L4k01 both librate with a period of $\sim$4--5~Myr, 
both are in the $\omega=270^\circ$ Kozai 
island\footnote{We do not believe there is any statistical significance to 
both Kozai objects being in the same $\omega$ island; the MPC sample has 
roughly equal numbers in each island.  Pluto itself is in the 90$^\circ$ 
island.},
and have libration amplitudes of $A_\omega$=30$^\circ$ and 50$^\circ$,
respectively.

With only two Kozai plutinos, the modeling we have done exceeds the
level of detail needed to deal with the detections, but we present our
efforts as a guide to the modeling that will be needed once characterized
samples grow.
We used the fourth-order averaged Hamiltonian given by \citet{WanHuang2007} 
to provide a reasonable approximate dynamics for the Kozai 
plutinos in the CFEPS survey simulator (see Appendix for details).
Kozai-librating plutinos have coupled oscillation of $\omega$ and $e$
(and hence $i$ because the product $\cos i \sqrt{1 - e^2}$ is constant,
 proportional to the angular momentum's z-component) 
that are determined by the value of $\cos i_{max}$ corresponding to the 
$e=0$ trajectory with the same angular momentum.
Looking at the full set of Kozai plutinos in the \cite{GladmanNomen2008} 
compilation, we found that using a set of Kozai trajectories 
corresponding to the $i_{max} = 23.5^\circ$ Kozai Hamiltonian with different
initial $e_{min}$ values provided a range 
of Kozai librations sufficient to model the current sample. 


Having this Kozai dynamics model, we proceeded to modify the L3 plutino
model by introducing the Kozai fraction $f_K$ parameter, which is the 
intrinsic fraction of the plutinos that are also librating in the Kozai 
resonance.
By running a one-parameter set of models, we find that an
intrinsic Kozai fraction of $f_K$=10\% gives the apparent CFEPS 
fraction of 8\%; that is, given the longitude coverage of the CFEPS,
there is a mild bias against the detection of Kozai librators.
This $f_K$=10\% fraction is similar to previous estimates
\citep{ChiangJordan2002,Nesvornyetal2000}.
Although not very constraining, our formal 95\% confidence upper limit is 
$f_K < 33\%$  
so many more plutinos from characterized surveys will be required
to accurately measure $f_K$.


\citet{TiscMal2009} point out that because the Kozai plutinos are 
somewhat more stable than the average plutino, the Kozai fraction should 
have slowly grown with time.  
Dynamical simulations which attempt to create the plutino orbital
structure must thus `erode' their populations to the modern epoch
and then state distributions of libration amplitude for both the
3:2 resonant argument and the Kozai libration amplitude, which may
be matched to future de-biased surveys. 
LSST may provide enough resonant TNO detections 
\citep{LSSTbook2009}
to use these distributions as diagnostics.

\subsection{The plutino size and eccentricity distribution}

\begin{figure}
\centering
\includegraphics[scale=1.1]{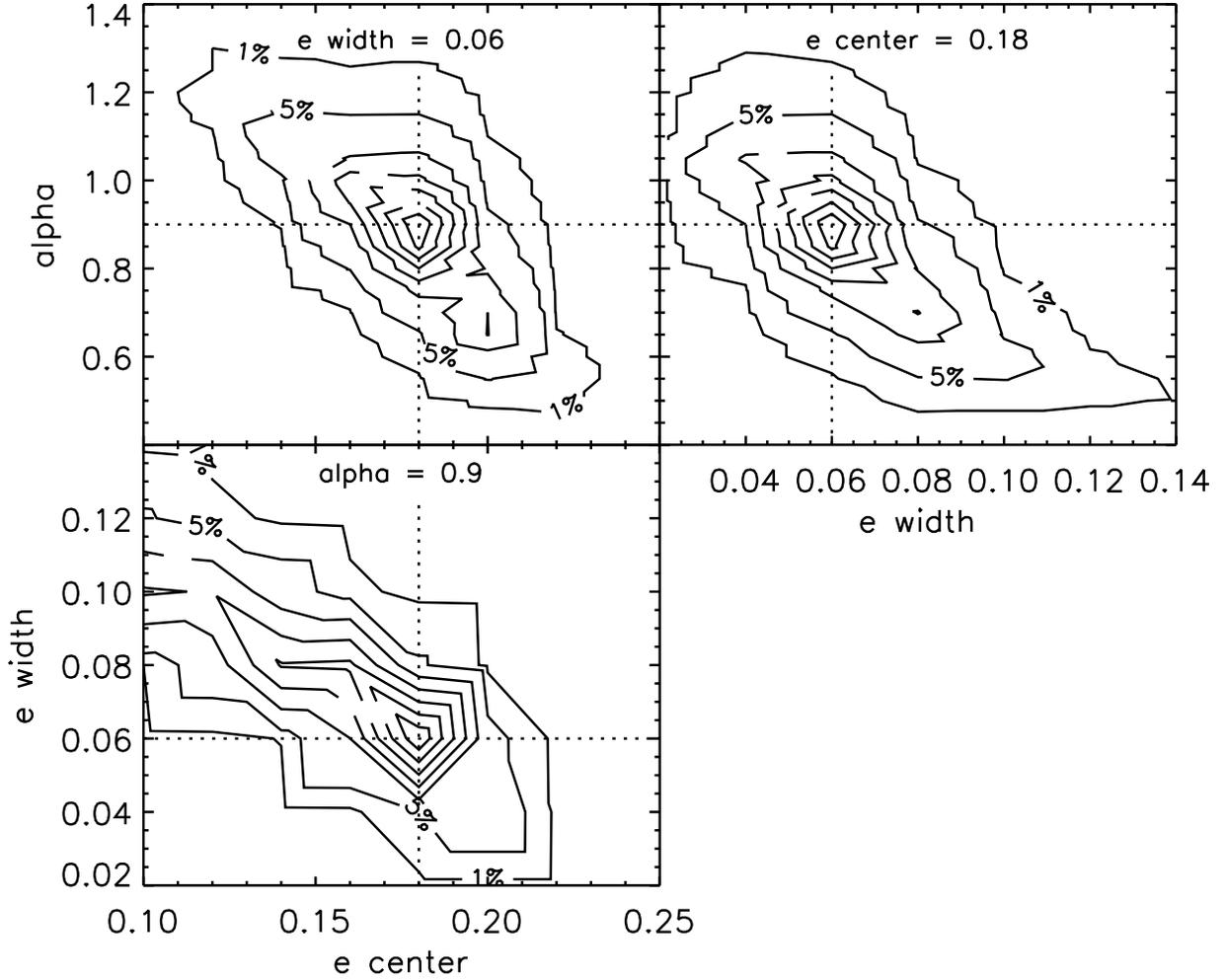}
\caption{Confidence regions in the plutinos orbital model parameter space.  
Three perpendicular slices through the $(\alpha,~e_c,~e_w)$ parameter 
space, showing  the regions interior to which none of the cumulative
distributions yield probabilities $<$5\% or $<$1\%.	
Note the coupling between the parameters; for example, smaller
values of $e_c$ are allowed only if the width $e_w$ and the 
$H$-magnitude distribution $\alpha$ both rise (or the detection 
distance distribution will fail as not being confined enough to 
small distances).
}
\label{fig:aewec32}
\end{figure}

With 24 detections, we can now independently measure the 
standard $H$-magnitude distribution slope $\alpha$ in the 
formulation $N(<H)\propto 10^{\alpha H}$ for the 
plutinos.
This is important because, as \citet{L3paper} showed, the distribution
of plutino detection distances is a sensitive function of the combination
of the $\alpha$, $e_c$, and $e_w$ parameters (where the latter two are 
the center and width of a gaussian $e$ distribution).
Detection biases favor finding larger-$e$ plutinos at 
small distances.
This is simply understood; when a small-body population has a steeply
increasing power-law size distribution,
any flux-limited survey is very strongly biased towards detecting the 
hordes of smaller objects that come above the flux limit only at
perihelion.
Because of these considerations, surveys really only measure the
slope of the size distribution which correspond to the $H$-magnitude
range for the population in question near perihelion; for CFEPS plutinos
this means that we constrain the value of $\alpha$ for the range
$H_g$=8--9; smaller plutinos are undetectable and larger ones are too rare 
to be statistically constrained.

We proceeded to run a very large grid of models covering the plausible
ranges of $\alpha$, $e_c$, and $e_w$, as preliminary explorations clearly
showed these parameters were correlated.
The results (Fig.~\ref{fig:aewec32}) give confidence regions
for our plutino model, where the figure shows cuts in three 
perpendicular planes through the best-matching model, with
$\alpha=0.9$ for all plutinos, and $e_c=0.18$ and $e_w=0.06$
for the non-Kozai component (however, these parameters remain
valid even if the Kozai sub-population's dynamics is ignored in the 
modeling).
For these experiments the inclination distribution and libration
amplitude are kept fixed (and experiments showed they are only weakly
coupled to the $\alpha$, $e_c$, and $e_w$ triad).
A model is rejected if at least one of the distance, eccentricity, or magnitude
distributions of the simulated detections disagree (via an Anderson-Darling statistical test)
with the real CFEPs detections.
We consider models outside the 5\% contour rejected.

As Fig.~\ref{fig:aewec32} shows, our 24-plutino sample is able to 
meaningfully constrain the properties of the plutino size and orbital
distributions.
The cumulative $e$, detection distance, and apparent $m_g$ distributions 
corresponding to 
our nominal model ($\alpha=0.90, e_c=0.18, e_w=0.06$) were shown in 
Fig.~\ref{fig:cumuplots32}.
As can be seen, there is the expected mild bias towards the detection of
higher-$e$ plutinos.
Much stronger is the remarkable bias seen in the (heliocentric)
discovery distance $d$ distribution; 22 of the 24 CFEPS plutino were detected 
with $d<a_{3:2}=39.4$~AU, even though any object on an eccentric orbit
spends more than half its time further than its semimajor axis.
As Fig.~\ref{fig:facedown32} shows, CFEPS covered a large range of ecliptic 
longitudes
and is thus extremely sensitive to the plutino distance distribution.
It is not surprising that the most distant CFEPS plutino (L5c11) is roughly
opposite to Neptune on the sky.
However, the preponderance of low-$d$ detections demands steeper 
slopes for the magnitude distribution and large median eccentricities
$e_c$.
The median plutino $e$ in the Minor Planet Center from the 
\citet{GladmanNomen2008} plutino compilation is 0.224; the hypothesis
that the true median intrinsic $e$ is this large or higher is ruled out at
$>$99\% confidence.

\begin{figure}
\includegraphics[angle=-90,scale=0.48]{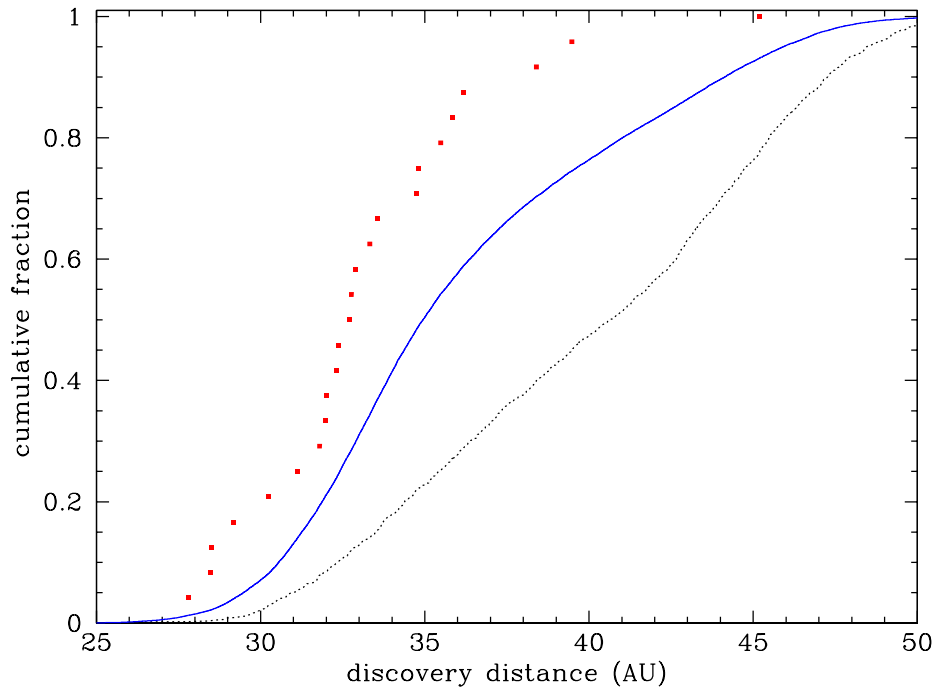}
\caption{The 
cumulative plutino distance at detection distribution for a model
with size distribution exponent $\alpha$=0.55.
Dotted black line is true heliocentric distance $d$ distribution, which
would be detection biased by the CFEPS survey to the solid blue
curve; red dots are the CFEPS plutino detections.
For such a flat size distribution too many large TNOs exist at 
great distance to be detected, which is inconsistent with the
concentration to small $d$ present in the CFEPS detections (this
model is rejected at more than 99\% confidence).
}
\label{fig:d_alpha55}
\end{figure}

This analysis demonstrates a correlation between the acceptable values
of $\alpha$, $e_c$, and $e_w$.
Somewhat shallower $H_g$-distributions ($\alpha$ down to 0.6) are allowed
within the 95\% confidence range, but 
such a size distribution requires an $e$ distribution peaked at larger 
values to maintain the dominance of small-$d$ detections.
While $\alpha$ down to 0.6 is not formally rejected by CFEPS,
slopes lower than this result in too large a fraction of distant 
detections.
Fig.~\ref{fig:d_alpha55}
illustrates how going to models beyond the 95\% confidence limit alters the
$d$ distribution dramatically.
Using $\alpha\simeq0.55$ and the best possible $e_c$ and $e_w$ values still 
results in a rejectable detection-distance distribution and, unlike our 
strong suspicion in \cite{L3paper}, we can now formally reject the 
suggestion in \cite{HahnMalhot2005}
that the plutino size distribution is as shallow as $\alpha=0.54$.
On the other end, $\alpha$=1.15 actually mildly improves the $d$ distribution
match, 
but such a model results in a $g$-magnitude distribution of the 
simulated detections being so strongly confined to magnitudes slightly 
brighter than 24 that this rejects the model at $>$95\% confidence.

The `least rejectable' model we have found has a size index
$\alpha$=0.9, corresponding to a diameter ($D$) distribution
with differential $dN/dD \propto D^{-5.5}$. 
Again, 
CFEPS measures this slope only in the $H_g$=8--9 range which dominates
the CFEPS plutino detections.
It is interesting to compare this to the $\alpha$=0.8 estimate from
\citet{L7paper}
for the classical main-belt hot population, measured for the 
$H_g=$7--8 range (the plutinos detections are dominated by physically smaller
objects than the more distant main-belt detections). 
The 0.1 difference between the two estimates is not significant, given
the uncertainties.
Due to the similarity in inclination and size distributions, our
working hypothesis is that the hot population and plutinos (and,
as we shall see below, the other resonant populations) share a 
common origin.

The uncertainty in $\alpha$ makes no significant difference to our
plutino population estimate.
If $\alpha$=0.8 (instead of 0.9) our estimate for the $H_g<9.16$
plutino population drops 
only $\sim$8\%,
a difference which is much smaller than the current 
population uncertainties (see below).

\subsection{Plutino libration amplitudes}
\label{sec:plutLibAmps}

Libration amplitudes of the 3:2 resonant argument\footnote{
Kozai plutinos still have their $\phi_{32}$ argument librate, with the 
argument of pericenter librating roughly two orders of magnitude slower 
than $\phi_{32}$.}
vary for CFEPS plutinos from $L_{32}=28^\circ$ (L4j11) to 
$125^\circ$ for L4m02.
Numerical simulations show that,
in the present planetary configuration, 
plutino libration amplitudes $L_{32}$ larger than about 125--130$^\circ$ are 
unstable over the age of the Solar System \citep{NesRoig2000,TiscMal2009}.
Any libration amplitudes $>130^\circ$ will be eroded away in the following 
4~Gyr of evolution, but most smaller-amplitude librators will be stable.
What cosmogonic processes set the distribution of the remaining
stable libration amplitudes?
\cite{LevStern1995pcb} show libration amplitude distributions generated
in a plutino population captured via gravitational scattering and then
damping into the 3:2.
\cite{ChiangJordan2002}
show different libration-amplitude distributions produced by sweep-up
capture, depending on Neptune's migration speed.

We first reconfirmed that a uniform $L_{32}$ distribution from
0--130$^\circ$ was rejected ($>$98\% confidence).
This test also showed that CFEPS has a mild bias towards detecting
plutinos with $L_{32}<100^\circ$ due to the longitude coverage.
Note that this bias is {\it not} generic to all TNO surveys; it 
depends strongly on the longitude coverage and depths of the survey; 
the $L_{32}$ panel of Fig.~\ref{fig:cumuplots32} shows that 
CFEPS over-detects plutinos in the $L_{32}$=50--100$^\circ$ range relative 
to their true intrinsic fraction.
However, the survey simulator allows us to remove this bias.
Compared to the L3 plutino model \citep{L3paper},
we are now able to meaningfully constrain the libration-amplitude
distribution.
The L3 model used a symmetric triangle probability distribution
motivated by the $L_{32}$ compilation in \citet{LykMuk2007}; 
that is, a probability that increases linearly
from $L_{32}$=0 to a peak at 65$^\circ$ and then decreases linearly
to $L_{32}=130^\circ$.
The L7 sample shows that this symmetric triangle is now a rejectable
representation of the true distribution, producing too many low-libration
amplitude plutinos.
We decided to modify the model by changing the low-$L_{32}$ start
of the linear distribution and its peak; the linear drop to the end of
the probability distribution was retained.
An end to the distribution just above the 125$^\circ$  amplitude of
L4m02 (which has the largest-known amplitude) is favored by
survey simulator analysis of the CFEPS detections.
We found that a start of the linear probability distribution at 
$L_{32}=20^\circ$ with a peak at $95^\circ$ provided the best 
`asymmetric triangle' probability distribution.
We tried expanding the range of libration amplitudes to different lower
and upper limits while holding the peak of the $L_{32}$
distribution constant at 95$^\circ$.  
The lower limits explored were 0 or 20$^\circ$, and the upper limits 
were 140, 150, 160, and 170$^\circ$.  
While none of these distributions were rejectable at 95\% confidence, 
they provided poorer matches to the CFEPS data than our 20 and 
130$^\circ$ nominal model for the lower and upper limits.

Only after arriving at this nominal model did we realize that the resulting
asymmetric triangle is very similar to the libration amplitude distribution 
shown in Fig.~6 of \citet{NesRoig2000}, which estimates the $L_{32}$
distribution for surviving particles in the main core of the resonance
after 4 Gyr of dynamical erosion, based on an assumed initial uniform
covering of resonant phase space.
We do not think that the de-biased CFEPS sample is able to constrain
fine details of the current (and thus initial) libration-amplitude 
distribution,
but it is clear that the mechanism which emplaced plutinos must be
capable of populating small libration amplitudes efficiently.

\section{The population of plutinos}

Comparison with other resonant populations is discussed in 
Sec.~\ref{sec:pops}, but we here put our plutino population estimate
in the context of  previous literature.
CFEPS is sensitive essentially all the way down to $H_g$=9.16 for
plutinos, which corresponds to the frequently used 100-km reference
diameter in the literature (for 5\% albedo).
The CFEPS estimate is
\begin{equation}
N_{\mathrm{plutinos}} (H_g < 9.16) = 13,000^{+6,000}_{-5,000} \; \; 
(95\% \; \mathrm{confidence}) \; .
\end{equation}

This can be compared to factor-of two estimate of 
1,400 from \citet{Truj2001stat},
which is the last published measurement independent of CFEPS, and the
previous CFEPS L3 \citep{L3paper} factor of two estimate 
of 6000 (scaled to $H_g<$9.16 utilizing the $\alpha=0.72$ slope, which 
now appears to be underestimated).
The L3 plutino estimate is consistent with our current estimate,
and remains discordant with \citet{Truj2001stat} for the same
reasons given in \citet{L3paper}.
Table~\ref{tab:pop} lists both the median $H_g<9.16$ estimate (which
we adopt as standard for all our absolute resonant population estimates,
being the limit to which CFEPS had high sensitivity) and an $H_g<8$
estimate because this value is the CFEPS sensitivity limit in the
classical belt, allowing comparison to that population. 
Due to the different size dependencies now being used, the 
\citet{L3paper} $H_g < 10$ estimate should be scaled to 
the $H_g<8$ limit by dividing
by $10^{2\alpha} = 10^{2(0.72)}\simeq30$ .

Given our current estimate \citep{L7paper} of the main classical belt
having 130,000 $H_g<9.16$ TNOs, the plutino population is thus
$\sim10$\% of the entire main classical-belt population at
the $9.16$ limit.  
Note that the L3 classical-belt estimate was only a restricted
portion of the main-belt phase space, and the L7 model now essentially
covers the entire non-resonant phase space from 40--47~AU.
It is important to stress that the plutino/classical population
ratio is $H$-mag dependent due to the steeper slope of the cold
component of the main classical belt.
Thus, for $H_g<$8.0 the plutino/main-belt ratio is 15\%, in agreement
with the estimate of \citet{L3paper}.

\section{The 5:2 resonance}

The dynamics of the 5:2 resonance are similar to that of the 3:2 in
that low libration-amplitude TNOs in the 5:2 come to perihelion at
a range of longitudes near $\pm90^\circ$ away from Neptune.
The first real 5:2 resonators were recognized by 
\cite{ChiangDES2003}.
As usual, the libration amplitude $L_{52}$ measures oscillations of the
resonant angle 
$(\phi_{52} = 5 \lambda - 2\lambda_N - 3\varpi)$ around a mean of 
180$^\circ$.
Thus, the detection biases are similar to plutinos, making population 
comparisons likely more robust.
Due to  their larger semimajor axis near 55.4~AU,
5:2 TNOs spend a large fraction of their orbital period further away
than even the most distant plutinos.
Although Fig.~\ref{fig:pretty} shows that at a given ecliptic longitude 
low-$L_{52}$ TNOs could be found at several different discrete
distances due to their phase behavior, an eccentric orbit still massively
biases the detections to be at the perihelia longitudes (constrained by
the libration amplitude $L_{52}$ of $\phi_{52}$).

The five CFEPS 5:2 objects all have remarkably-high eccentricities
(in the narrow range 0.38--0.42), inclinations from 2--23$^\circ$,
and $L_{52}$=44--91$^\circ$.
Because the MPC has 5:2 resonators with $e<0.38$, we think this concentration
for $e\simeq$0.4 is a statistical fluke; a similar situation occurred with
the plutino discovery distances in \citet{L3paper} which disappeared in the
current larger sample.
Some 5:2 resonators with well-determined orbits in the MPC sample 
have eccentricities below $e\sim$0.3.
With only 5 CFEPS detections we cannot place strong constraints on
the internal orbital distribution, so we proceeded to build a model
with a similar level of detail as the \citet{L3paper} plutino
model.
Luckily, the 5:2 lacks a Kozai sub-component (no known TNO librates in
the 5:2, and we are unaware of any theoretical prediction indicating
a non-negligible phase-space for Kozai inside the 5:2).

The inclination distribution is consistent with being the same as
that for the plutinos; Table~\ref{tab:pop} lists the `least rejectable'
value of $\sigma=14^\circ$, but the large uncertainties mean identical
inclination distributions for 3:2 and 5:2 TNOs is a plausible
hypothesis, which we thus adopt..
As for the plutinos, we ran a 3-dimensional 
$(\alpha,e_c,e_w)$ grid to set confidence intervals on these parameters.
Unsurprisingly, this analysis did not meaningfully constrain $\alpha$ (which
allowed the range 0.4--1.2 at 95\% confidence, with a broad peak around
$\alpha$ $\sim$ 0.9).
We thus chose to use the plutino-determined value of $\alpha$=0.9 for 
the 5:2 and all other $H$-magnitude distributions for resonant populations 
which CFEPS had a detection.

\begin{figure}[h]
\centering
\includegraphics[scale=0.7]{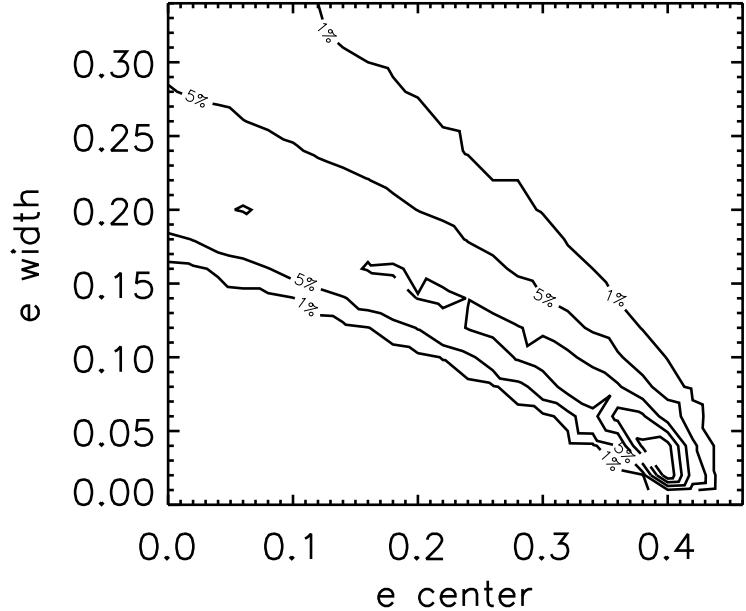}
\caption{Confidence regions for the 5:2 parameter space, for $\alpha$=0.9, 
showing the eccentricity distribution's range of acceptable 
$(e_c,~e_w)$ parameters.
Although CFEPS most favours a strongly-peaked distribution near $e=0.4$,
a region of lower $e_c$ centers with larger widths is acceptable
inside the 5\% limit. Even a Gaussian centered on $e_c$=0 cannot yet be
formally rejected due to the strong detection bias towards 
larger $e$.
}
\label{fig:ewec52}
\end{figure}
 
The detected eccentricity distribution for 5:2 resonators is 
obviously different than for the plutinos; 
eccentricities up to $e\simeq 0.4$ exist, corresponding to 
$q\simeq$30~AU.
The prevalence of orbits with perihelion at Neptune might be taken 
as firm evidence that this population was emplaced by scattering, but 
the detection biases also favour low $q$, so perhaps there are abundant
low-$e$ 5:2 resonators that make up only a small fraction of a
detectable sample.

We performed a similar grid search for acceptable parameters of an
eccentricity distributions with a gaussian center $e_c$ and
width $e_w$.
Fig.~\ref{fig:ewec52} shows that the most-favoured model is indeed
a narrow peak centered near $e_c=0.4$.
Like the plutinos, there is a coupling between acceptable values 
of the distribution's width and center.
It is possible that $e_c$ is much lower and the width $e_w$ higher.
We find this result to be generic for all the Kuiper Belt resonances;
it simply results from extreme bias towards detecting the abundant
small TNOs near the perihelia of high-$e$ orbits. 
Compared to the 3:2, the quality of fit of lower $e_c$/higher $e_w$
pairs is not in as great a contrast with the quality of the most
favoured case. 
We thus elected to base our nominal 5:2 model and population
estimate (Table~\ref{tab:pop}) on ($e_c,e_w$)=(0.3,0.1) 
(along the ridge) instead of the absolute peak where 
($e_c,e_w$)=(0.4,0.04);
the latter case has a 40\% smaller population but cannot
be correct given that two 5:2 resonators (2005 SD$_{278}$ and 
84522) exist in the MPC sample with $e<0.3$.
If the real 5:2 population has many even-lower eccentricity orbits,
the population will be somewhat larger than our nominal estimate; for 
example, even the rather extreme case of ($e_c,e_w$)=(0.14,0.18)  
yields a population 40\% larger than our nominal estimate.

The nominal 5:2 model produces a population estimate of 12,000 5:2
resonators with $H_g<9.16$
(Table~\ref{tab:pop}).
Although the 95\% confidence limits range from 4,000-27,000,
the favoured population is, perhaps surprisingly, essentially equal to 
that of the plutinos.
This is an unexpected result, as it indicates that the 
detection bias against 5:2 is roughly a factor of 5 stronger,
due to the larger values of $a$ and $e$, both of which result
in the population being much less detectable than the 
plutinos.
We will return to the cosmogonic implications of this in 
Sec.~\ref{sec:pops},
having compiled population estimates for other resonances.

The ability to capture TNOs into the 5:2 via either sweeping
up a pre-existing belt or capturing scattering TNOs into the
resonance was discussed by \citet{ChiangDES2003}.
These authors showed that although resonance sweeping could capture 
into the 5:2, the observed orbital-element distribution and the 
apparent 5:2/2:1 detection ratio could only be explained if the 
resonances captured objects with a pre-excited $e$ and $i$ 
distribution.
Creating most 5:2 TNOs by `resonance sticking' out of a disk
of TNOs scattering off Neptune (in the current planetary
configuration) was argued to be untenable.

The \citet{Levetal2008} scenario of having the resonant populations
trapped during a phase of outward migration can produce lower-$e$
and lower-$L_{52}$ TNOs after Neptune's eccentricity is damped,
and in one simulation produced a concentration of $e\sim0.4$ 
resonators (although this simulation fails to produce other
needed constraints like the inclination distribution).
This scenario is promising as a general way to trap resonant
populations out of an already-scattering population. 
In their comparison with 5:2 resonators from the most successful 
\citet{Levetal2008} run (for that resonance) with the MPC sample, the 
total range of $e$, $i$, and $L_{52}$
almost span the values of known MPC TNOs, but their comparison
was not corrected for observational biases which favour
low-$i$ and high-$e$ detections, which means that the run
produces a simulated 5:2 population that has too many large-$e$
and low-$i$ orbits.

The following two sections deal with the $n$:3 and $n$:4 
resonances.
Some readers may wish to skip forward to Section~\ref{n:1}'s
discussion of the $n$:1 resonances and cosmogonic significance,
especially on a first reading.

\section{The n:3 resonances}

The $n$:3 resonances have 3 different ecliptic-longitude centers 
(Fig.\ref{fig:pretty} shows examples)
at which objects are currently coming to perihelion: one is
opposite Neptune and the others are 60 degrees ahead and behind.
A resonant-argument libration of amplitude $L$ then results in
an angular deviation of $L/3$ on the sky of the perihelion-longitude
location relative to these three centers, over the course of a 
full cycle of the resonant argument.

CFEPS has detections in the 4:3 (4 TNOs), 5:3 (6 TNOs) and 7:3 (2 TNOs).
The detection biases for these $n:3$ resonances are similar in the
sense that the CFEPS block locations will not favour one of these
three over the other unless they have different libration amplitude
distributions, for which we see no evidence.
However, the smaller semimajor axis objects will be favoured
due to fraction of time spent in the detection volume.

With a few detections per resonance, we have not attempted to model 
the internal structure of the resonances, but have made a simple
generalization of the $n:2$ resonances.
We keep the same non-symmetric triangle for the libration amplitude
distribution as for the 3:2 (with no amplitudes above 130 degrees).
The eccentricity width $e_w$ is also retained, but the central value $e_c$
was moved to correspond to $q\simeq$33--35~AU. 
This is consistent with an idea that the resonant objects were largely
trapped from a primordial Neptune-coupled population, but is not
required by our data.
Lower values of $e_c$ are allowed in the same sense as the discussion of 
the 5:2 resonance; detection biases sufficiently favour the high-$e$ TNOs
that lower-$e_c$/higher-$e_w$ pairs are permissible (which would slightly
raise the population estimates).
We retained $\alpha$=0.9 for these resonances.

We did not retain the $\sigma\simeq 16^\circ$ inclination width from
the plutinos, as we found all three $n:3$ resonances favoured somewhat
lower inclination widths (although the 95\% confidence regions allow
$\sigma = 16^{\circ}$).
Table~\ref{tab:pop} lists the favored $\sigma$ width for each resonance
(where the 7:3 is extremely uncertain, so $\sigma=10^\circ$ was used)
along with the population estimates for $H_g<$9.16 and $<$8.0.
The 4:3 population must be small ($<$2,000 with $H_g<9.16$, at
95\% confidence). 
Although we have five 5:3 resonators and two 7:3 resonators,
the bias against the 7:3 TNOs (which have larger $a$ and $e$ values)
results in the true 5:3 and 7:3 being roughly equal 
(at $\sim 5$ times the 4:3 population).

\subsection{The 4:3 resonance}

The 4:3 resonance at $a\simeq36.5$~AU was studied by 
\citet{NesRoig2001}, who showed
that a resonance amplitude distribution like the 3:2 (of an asymmetric
triangle with peak near $L_{43}$=80--90$^\circ$) represented those 4:3
TNOs that survive over the age of the Solar System.
Although not heavily explored, these authors provide some evidence that
the stability of the resonance is not a strong function of inclination;
if this is also true for the 5:3 and 7:3 resonance then confirmation
of a colder inclination distribution for TNOs currently in the $n:3$
resonances would require a cosmogonic explanation (as opposed to being
due to later dynamical depletion).
\citet{NesRoig2001} calculate that, under a simple scenario of excitation 
of a primordial belt with initial surface density dropping as $r^{-2}$,
the number of 4:3 resonators should be 0.77 that of the 3:2 population,
whereas our estimate is 0.06, with 0.77 excluded at more than 95\%
confidence. We thus confirm that this scenario is excluded, and any 
Kuiper Belt structure-formation scenario must result in a very 
weakly-populated 4:3 resonance in the present epoch.

\subsection{The 5:3 resonance}

The 5:3 resonance at $a\simeq42.3$~AU
has the curious attribute that it is almost precisely at
the lower semimajor-axis limit of the low-inclination component in the
main part of the classical belt.
The instability for non-resonant TNOs is due to the $\nu_8$ secular 
resonance which removes low-$i$ TNOs just interior to $a$=42~AU.
The faster precession caused by the resonant argument for TNOs inside
the 5:3 shields its members from the $\nu_8$'s effects, so the proximity
of the 5:3 and beginning of the low-$i$ classical belt
is likely just a coincidence, and not of cosmogonic 
significance\footnote{Some mean-motion resonance can always be found
close to any given point in the main Kuiper Belt...}.

\citet{LykMuk2007} and \citet{GladmanNomen2008} list 11 
TNOs\footnote{There is a typo in Table 2 of Gladman {\it et al.} (2008) 
in the 5:3 entry for K03UT2S = 2003 US$_{292}$, whose unpacked designation is
mistakenly given as 2003 US$_{96}$. After 2008, this TNO was 
numbered 143751.}  
from the MPC as librating in the 5:3.  
CFEPS detected six 5:3 resonators, two of which were discovered before
2003 (Table~\ref{tab:elems2}).

\citet{MelitaBrunini2000} performed a numerical study of 5:3
resonators, showing that the interior of the resonance does not
contain a simply-connected stable region, and that lower-$e$ orbits
appeared more stable in a frequency-map analysis; comparison with real
objects was difficult as there was only one 5:3 resonator (1999 JS)
at the time; the objects plotted in Fig.~3 of \citet{MelitaBrunini2000}
with $e<0.15$ are non-resonant.
\citet{LykMuk2007} also explored the 5:3 numerically and found that
particles surviving the age of the Solar System were mostly concentrated
in the region $0.09 < e < 0.27$ and $i<20^\circ$, which is indeed 
the range occupied by the known 5:3 TNOs.

We note that the 5:3 eccentricities are much higher than 
the classical objects in surrounding semimajor axes (also obvious for the 
7:4).
It remains uncertain if this is because these resonant TNOs were 
captured from a lower-$e$ population and pumped to higher $e$ by 
migration, or rather if
both resonant and non-resonant objects existed with $e$ up to 0.25
and then nearby classical object were eroded away over the age of 
the Solar System.  
The former scenario seems disfavored when considering distant
resonances like the 7:3, which do not appear to have the low-$e$
members they might be expected from a sweep-up scenario into a 
pre-existing belt (although the selection bias against their
detection is strong).

At the other extreme, the {\it absence} of 5:3 TNOs with $e<0.10$ might
also be seen to argue against `sweep up' migration (because low-$e$ 
5:3 objects could be swept up during the final stages from the classical 
belt).
On the other hand, there is detection bias against the discovery of the
lowest-$e$ members and the \citet{LykMuk2007} integrations show 
that such low-$e$ 5:3 resonators can leak out into the surrounding
classical belt.

\subsection{The 7:3 resonance}

The 7:3 mean-motion resonance (with $a\simeq 53.0$~AU) is little
discussed in the literature due to being beyond the 2:1 resonance
and being 4$^{th}$ order (and thus nominally weaker).
\citet{LykMuk2007} and \citet{GladmanNomen2008} each list 3 TNOs
in the 7:3 resonance; only 2 of the TNOs were shared (2002 GX$_{32}$ = 95625 
and CFEPS L3y07 = 131696)
at the time, but due to improved orbital information both 
2004 DJ$_{71}$ and 1999 CV$_{118}$, and
perhaps 1999HW$_{11}$, are also likely 7:3 resonators.
Additionally,
the CFEPS object L5c19PD is a re-discovery of the lost object
2002 CZ$_{248}$, whose orbit based on a 1-month arc was given to
be $a\simeq 56.6$~AU and the ephemeris was about 0.5 degrees away
from the prediction by the time of our 2005 discovery; CFEPS
tracked the object a year before being able to establish the linkage
to the short arc from 3 years earlier.

With only two CFEPS detections, we are unable to strongly constrain the
parameters that govern the internal orbital distribution.  
We find (Table~\ref{tab:pop}) that a model with inclination width
like the other n:3 resonances of $\sigma=10^\circ$ 
and eccentricity width $e_w$=0.6 works acceptably as long as 
the $e$ distribution is centered on $e_c$=0.30 so that perihelia 
in the $q=30$--35~AU range are allowed. 
As before, lower $e_c$ coupled to larger $e_w$ cannot be 
excluded.  This yields population estimate of 4,000 7:3 resonators
to factor of three accuracy at 95\% confidence, about a factor of three 
below the 3:2 and 5:2 populations.

\section{The n:4 resonances}

The 5:4 and 7:4 resonances are little discussed in the literature, despite 
them both being populated.
The \citet{GladmanNomen2008} compilation lists nineteen 7:4 librators and 
three 5:4 resonators.
The resonant argument forces pericenters to be in bands
centered on $\pm45^\circ$ and $\pm135^\circ$ away from Neptune.
Due to the proximity of these locations to the galactic plane,
observational surveys have probably not covered these regions 
as well as they cover the pericenter longitudes of the $n:2$
resonances.

\subsection{The 5:4 resonance}

CFEPS has only one 5:4 resonator and the DES survey \citep{ElliotDES2005} 
a second, bringing the current total to five known objects.
With $a\simeq$35~AU, the 5:4 is the closest (in semimajor axis)
exterior mean-motion resonance to Neptune that is known to be 
populated, but the proximity to Neptune makes the stable phase space
restricted.
\cite{Malhot1996} showed how the zone of stable libration amplitudes
shrinks rapidly with increasing $e$; all known 5:4 librators have
$e$ in the range 0.07-0.1.  
With one CFEPS detection we provide an estimated population of
$N(H_g < 9.16)\sim$160, with factor of five 95\% confidence
limits.
Despite its relative uncertainty, it is clear that the 5:4 population 
is at least an order of magnitude less populated than 
the 3:2 or 5:2.

\subsection{The 7:4 resonance}

The dynamics of the 7:4 resonance at $a\simeq 43.7$~AU were discussed by 
\citet{LykMuk7to4},
who showed that the maximum stable amplitudes drops as eccentricities
rise.
These authors noted that the most dynamically-stable part of the resonance
($e$=0.25--0.30 with $i$=0--5$^\circ$) appears unpopulated, despite
it being easier to find TNOs with these eccentricities than the 
lower eccentricities of the known 7:4 resonators
(the largest-$e$ CFEPS 7:4 has $e$=0.21, while the MPC's orbit for
2003 QX$_{91}$ has $e$=0.25).
Dynamical simulations \citep{HahnMalhot2005,Levetal2008,YehChang2009} rarely
show occupation of the $e>0.25$ region, so the lack
of $e>0.25$ 7:4 TNOs seems in line with model results that this region 
was not populated during the Kuiper-belt sculpting process.

Examinations of the dynamical `clones' of the nominal classifications show
that the phase space of the resonance is extremely complex.
Even relatively long-arc orbits show great variation in libration
amplitude amongst the clones, and thus our tabulated libration 
amplitudes are only accurate to a factor of 2.
A striking aspect of the CFEPS 7:4 detections is their preferentially-small
inclinations.
When fitting a $\sin(i)$ times a gaussian distribution, we reach the
same conclusion as 
\citet{Gulbis2010}
that the acceptable $\sigma$ widths are considerably colder than for other
Kuiper Belt sub-populations.
Our 95\% confidence range for the inclination width is 2.5--14$^\circ$, 
with $5^\circ$ being favored, in good agreement with the 
Gulbis {\it et al.} result.
\citet{LykMuk7to4} had already shown that 7:4 resonators with $i>10^\circ$ 
are much less likely to survive the age of the Solar System; thus the
colder inclination distribution cannot be taken to be a direct signature 
of the trapping process, although the preference for $e<0.25$ may be 
such a test.

\section{The n:1 resonances} 
\label{n:1}

The $n$:1 resonances require more modeling care  because of the presence of 
symmetric and asymmetric libration islands
(see \cite{Beauge1994} and citations to it).
That is, instead of the resonant argument oscillating symmetrically
around 180$^\circ$, there are three possible modes.  
The symmetric mode is centered on 180$^\circ$ but, unlike for the
resonances discussed earlier, there is a lower limit for the 
symmetric libration amplitude because the asymmetric islands
occupy the phase space where low-amplitude libration occur.
The asymmetric librators have libration centers that depend on
the TNO's orbital elements (especially its $e$) and have an upper
bound to their libration amplitudes \citep{Malhot1996}.
Detailed modeling of the $n$:1 resonances would require much more
information than the small number of CFEPS objects provide.
We have thus chosen to use orbital models motivated by analytic 
studies of the resonances, where our adjustable parameters are
confined only to the inclination distribution and the fraction $f_s$
of the TNOs that are in the symmetric mode.
The population estimates thus have some dependence on the accuracy
of the analytic studies.
Although some $n$:1 librators also show evidence of simultaneously
being in the Kozai resonance \citep{LykMuk2007}, we simply do not
have the numbers of detections to warrant modelling this as an
additional sub-component; as for the the plutinos we expect that 
the population estimates are only very weakly ($<$10\%) dependent
on the presence or absence of the Kozai sub-component.

\subsection{The Twotinos}

The name `twotino' has been given to 2:1 resonant librators.
Much has been made in the past of the population ratio of
plutinos to twotinos, because this may be diagnostic of migration
models \citep[{\it eg}.,][]{Malhot1995,JLT96,ChiangJordan2002}.
An important goal for us has thus been to provide an estimate
of the twotino population ratio to both the 3:2 and 5:2 resonances 
(which we discuss in section \ref{sec:pops}).
In addition, \citet{ChiangJordan2002} showed that Neptune's migration rate
could affect the population ratios of one asymmetric island to the other.

The CFEPS sample provides 5 characterized twotinos 
(Table~\ref{tab:elems2}).
Another twotino detected in the survey, 
U7a08
\citep{L7paper},
is associated with the symmetric island but 
U7a08 
is excluded from this resonant study because its faintness puts it 
below the 40\% detection efficiency threshold which CFEPS felt it could 
reliably debias (``U'' means un-characterized).
Unfortunately this is the largest-inclination twotino ($i=7.0^\circ$) in
our sample.

We elected to use a gaussian inclination width of 9$^\circ$, which allows 
the CFEPS survey simulator to provide a large fraction of $i<7^\circ$
detections, while simultaneously allowing the existence of larger-$i$
twotinos known in the MPC 
sample\footnote{The
 only secure twotino with $i>15^\circ$ is 130391 = 2008 JG$_{81}$, with 
$i=23.5^\circ$,  which appears to be a symmetric 
librator}.
We find that the inclination distribution of the twotinos
{\it must} (at $>$95\% confidence) be colder than for the 3:2
and 5:2 resonances.
The lack of large-$i$ 2:1 librators in CFEPS is not statistically
alarming, especially when one considers that one does not expect the
inclination distribution today to be gaussian:
\citet{NesRoig2001} and \citet{TiscMal2009} show
that long-term dynamical stability of the 2:1 is inclination dependent,
with inclinations above 15$^\circ$ being more unstable, especially 
for symmetric librators.
Thus the colder inclination width does not necessarily provide cosmogonic 
information. 
We have verified that changing the inclination width by a factor of
two generates only a factor of two variation in the population estimate, 

Three of the four characterized asymmetric CFEPS twotinos occupy the 
island with $<\phi_{21}> \simeq 290^\circ$ (sometimes called 
the `trailing' island because the perihelion longitudes are `behind'
Neptune's ecliptic longitude) 
while the fourth occupies the leading asymmetric island.
We thus have an apparent (biased) measure of the `leading fraction'
$f_L^{biased}$ of 0.25.
This is an interesting contrast to
\citet{ChiangJordan2002}
who reported that all of the twotinos from the DES at that time inhabited 
the leading asymmetric island, and \citet{Ruth2005} discuss the 
apparent leading/trailing ratio of 7/2 at that time,
or $f_L^{biased}$ = 7/9 = 0.78.
It is clear that our trailing preponderance is due to the 
depth of the CFEPS L4j, L4k, L5i, and L5j blocks
\citep{L7paper}
which are well placed to find trailing twotinos, while the CFEPS
coverage of the longitude where leading asymmetric twotinos come
to perihelion is sparse (the Ls3s block was not especially deep).
We used the CFEPS survey simulator to show that on average one-third of 
asymmetric twotinos detected by CFEPS would be in the leading island 
($f_L^{biased}=0.33$) due to our block depth and placement relative to 
galactic plane, even if the true population was equally distributed 
($f_L$=0.5) between the two islands.
We hypothesize that the DES survey simply had had the opposite selection 
effect.
\citet{Ruth2005} suggested calibrating the observational selection
effects by using the 3:2 ratio, but the galactic plane confusion
is not the same for the two resonances; due to Neptune's position, 
`trailing' plutinos are not
as confused by the galactic plane as trailing asymmetric twotinos.
Therefore precise measurement of a population asymmetry demands an
absolutely-calibrated survey with well-understood detection efficiency
differences for the two relevant portions of sky.
To illustrate what limits can be set on the true value of 
$f_L$ using the CFEPS calibration, we asked the question:
How large would $f_L$ have to be before 95\% of the time CFEPS
would find 2 or more leading detections (and thus rule out this value
of $f_L$)?
The CFEPS calibration demands $f_L < 0.85$ at 95\% confidence.
For a lower bound, $f_L>0.03$ is required (95\% confidence) to
allow the existence of at least one leading twotino detection
in CFEPS.
The 67\% confidence range is $0.35 < f_L < 0.64$ but we prefer to
use the 95\% range of $f_L$=0.03--0.85 for the fraction of all
asymmetric twotinos in the leading island. 
The $f_L>$0.03 limit only requires that the trailing/leading ratio
be less than 30, to be compared with ratios up to $\sim10$ found
by \citet{Ruth2005} in simulations of asymmetric capture during
Neptune migration.
This weak observational constraint does not yet provide interesting rejection
of cosmogonic theories, but a factor of several more twotinos in
well-calibrated surveys has the potential to do so.

The symmetric librator K02O12=2002 PU$_{170}$ has libration amplitude
$L_{21}=154\pm4^\circ$; over a full libration cycle its perihelion
longitude can thus be found anywhere on the sky not within $\simeq 25^\circ$
of Neptune.
The excluded CFEPS discovery U7a08 (not characterized due to its faintness)
is an alternating `three-timing' 2:1 object, meaning that
during numerical evolution forward in time, its resonant argument
switches between symmetric and asymmetric modes.
This commonly-seen behaviour
\citep{ChiangJordan2002}
does not invalidate a parametrization of the 2:1 as having a 
`symmetric fraction' $f_s$ because it is reasonable to assume that
this fraction is maintained in steady state.

With only five characterized 2:1 CFEPS detections, our orbital 
distribution is based on abundant theoretical understanding of the
resonance's dynamics, rather than an empirical model fit to our
detections (which will have too many parameters to be constrained by our 
5 detections).  
Instead, the range of libration centers, amplitudes and eccentricities
(and correlations between them) are provided from analytic understanding
and numerical explorations of the resonance (see Appendix).
This model provides a non-rejectable match, leaving as the only
remaining adjustable parameter the unknown fraction $f_S$ that the
symmetric librators make up of the twotino population.

The symmetric libration fraction is poorly measured.
A few such objects are known \citep{LykMuk2007}, but again because the
selection effects are very different for symmetric versus asymmetric
librators only a survey with well-characterized sky coverage can
provide an estimate. 
With only 1 in 5 characterized detection in CFEPS being symmetric, 
we can only weakly constrain $f_S$.
Because the fraction of detected twotinos which are symmetric will 
depend on a survey's longitude coverage, we can only determine it
for our own survey;
we did this by running a large suite of models to determine the
detected fraction of symmetric detections as function of the intrinsic
value and find that our 20\% apparent fraction implies $f_S\simeq 0.3$,
which we adopt.
The remaining 70\% of the twotinos are equally divided among the
two asymmetric islands.
Luckily our population estimate is only a weak function of $f_S$; we
determined that even if $f_S$ were increased to 0.75 the total twotino
population estimate rises only 25\% (again, this result will not be 
identical for a survey with different sky coverage).

We find a plutino/twotino ratio to be $\sim$3--4, similar to the
ratio estimated by \citet{ChiangJordan2002}.
An important new result from CFEPS is the fact that the twotinos
are less numerous than 5:2 librators, which will be discussed
in Sec.~\ref{sec:pops}.

\subsection{The 3:1 resonance}

CFEPS detected two TNOs in the 3:1 resonance, one of which (U5j01PD) was 
below the 40\% detection efficiency threshold.
With these statistics we are unable to explore details of the 
TNO distribution
inside the resonance's structure; instead we provide a population 
estimate, which is likely only accurate to order-of magnitude.
The dynamics allows both symmetric and asymmetric
librators; \citet{Malhot1996} shows the 3:1's structure. 
Due to lack of constraint, we retain the symmetric fraction $f_S=0.3$ 
used for the twotinos.
We use an orbital-element distribution inside the resonance essentially 
the same as for the 2:1, excepting that the model eccentricities extend up 
to that necessary to reach $q\simeq 30$~AU.

\citet{ChiangDES2003} and \citet{HahnMalhot2005} demonstrated 
that 3:1 librators could be produced in an outward migration scenario 
into a initially warm ($e\sim0.1$) pre-existing belt.
In both simulations the initial disk extends to at least 55~AU from
the Sun, although it is not clear where the warm disk must extend to in order
to enable 3:1 trapping.
\citet{Levetal2008} do not provide information on the 3:1 (or more distant) resonances,
confining their discussion to $a<60$~AU.

The 3:1 librator L5j01PD  = 2003 LG$_7$ = 136120
(Table 4 of \citet{L7paper})
was discovered by the DES survey in 2003 \citep{ElliotDES2005} 
and independently re-discovered by CFEPS in 2005.
Despite observations in each and every of 5 sequential oppositions from
2003--2007, we are unable to securely determine if the object is a 
symmetric or asymmetric librator, although the symmetric case is favoured.
\citet{LykMuk2007} classified the object as symmetric using
the DES data from 2003 to 2006 inclusive (with amplitude $\approx160^\circ$
for the best-fit orbit) but we find that asymmetric libration is still 
allowed for orbits consistent with the astrometry.
This serves as another example of the need for abundant high-precision
astrometry to determine the details of the resonance dynamics.
Because this TNO has flux resulting in a detection efficiency below the
40\% limit in the L5j block, we do not use it in our population model.

The characterized TNO L4v08, with similar 5-opposition coverage, 
may also be either a symmetric or asymmetric 3:1 librator, with the 
former slightly favored.
The 2-night 2004 discovery of L4v08 was already in the MPC astrometric 
database with designation 2004 VU$_{130}$, with an orbit putting it
at the $d$=49~AU aphelion of an $a$=43.9 AU classical-belt orbit.
Note that L4v08 happens to be the most distant CFEPS resonant TNO at 
$d=49.7$~AU;  L5j01PD is at $d=33$~AU.
Because the 3:1 population must extend to $d\sim90$~AU, given their
$e$=0.4--0.5 range, the fact that in both cases $d\ll a$ again illustrates the 
extreme pericenter detection bias caused by the eccentricities and steep size
distribution.

Using a 3:1 model similar to the 2:1 model, our population
estimate is 4,000 3:1 TNOs with $H_g < 9.16$, with factor of
3 error bars at 95\% confidence.  
The resulting debiased CFEPS 2:1/3:1 ratio estimate of $\sim$1 is not 
statistically distinguishable from the $\simeq3.5$ estimate in the 
50-Myr migration simulation of \cite{ChiangDES2003}, given our 
uncertainties at 95\% confidence.
However, the Chiang {\it et al.} simulation does not `erode' the
surviving resonant populations (given 50 Myr after migration start) for 
the age of the Solar system, which could change the ratio.
\cite{HahnMalhot2005} show (see their Fig.~6) that 
for their model's emplaced populations the 2:1/3:1 ratio does not 
change much during erosion even if 
both populations drop mildly over 4 Gyr; however their 2:1/3:1 ratio is 
$\sim10$, which is rejectable at $>$95\% confidence.
Both models plausibly demonstrate the production of 3:1 TNOs with
$e>0.4$ and $i$ up to 20$^\circ$.

\subsection{The 5:1 resonance}

Our sole 5:1 TNO (L3y02 = 2003 YQ$_{179}$) was provisionally classified as 
a detached object by \citet{GladmanNomen2008}
but flagged as being insecure and quite possibly resonant in the 5:1
(despite already having a 3-opposition orbit in 2008).
Further tracking observations by our team have resulted in the
now-improved orbit being (insecurely) classified as a resonant 5:1 orbit.
The high-order resonances are so `thin' in phase space, that
we postulate other `detached' TNOs are actually in high-order
(meaning $j-k$ is large) mean-motion resonances as well.
In this case the largest-$a$ orbit consistent with the
astrometry is just outside the resonance; however, we are essentially
sure that this object is the first identified 5:1 librator.

We note that the detached object 1999 CF$_{119}$,
discovered by \citet{Truj2001stat}, has a semimajor axis
$\sim0.3$~AU beyond the 5:1 resonance, and the \citet{GladmanNomen2008} 
analysis indicates the lowest-$a$ plausible orbits are just barely 
beyond the resonant semimajor axis.  
A small systematic error in one opposition of the 4-opposition orbit
might suffice to remove the nominal orbit from the resonance; we thus
suggest additional observations.

With $a=88.38$~AU, $e$=0.579, and $i=20.1^\circ$, the detection biases 
against TNOs like L3y02 are extreme.
We used a 5:1 model similar to the 3:1, with asymmetric and symmetric 
($f_S=0.30$) 
librators, and an inclination width $\sigma=10^\circ$.
Using the single detection, we estimate 8,000 TNOs with $H_g<9.16$ in the
5:1 resonance, an estimate which is only good to a factor of five
given our lack of knowledge of the inclination distribution.
In particular, if the inclination distribution is considerably hotter
than the $\sigma=10^\circ$ value we have taken from the 2:1 (which seems
likely given that L3y02 has an inclination twice that value), then the
population estimate will rise.
Even at the nominal $i$-width, the 95\% confidence limits permit this
resonance to actually be the most populated of all trans-neptunian
space.


\section{The Neptune Trojans}

The first Neptune trojan was identified by \citet{ChiangDES2003}, 
and only $\simeq$7--8 are currently known \citep{ShepTruj2010Sci,Horner2012}.
The CFEPS survey did not discover a single Neptune 
trojan\footnote{Although the MPC currently lists L4k09 = 2004 KV18
as a L5 trojan
\citep{Horner2012}, 
the eccentricity of 0.184 is larger than numerically-determined 
stability limits \citep{NesDones2002}.
Although `near' the L5 cloud,
the \citet{GladmanNomen2008} analysis shows the object scatters
heavily on time scales $<$10~Myr and thus \citet{L7paper} reported
L4k09 as a scattering TNO, even if on a very short time scale it
may be temporarily near the L5 state.
Near-Earth asteroids exhibit similar temporary co-orbital behaviour 
\citep{MoraisMorby2002}.
}.
As the survey ran, we were very aware of the possibility of detecting
1:1 resonators, and confirm that this has nothing to do with possible detection 
biases in the survey.
The pericenter longitudes of many of the known resonances overlap with
the longitudes where Neptune trojans would spend their time, and CFEPS
found resonant and scattering TNOs at distances even closer than those
which Neptune trojans would approach; maximum eccentricities of the
known trojan sample \citep{ShepTruj2010Sci}
of $e\sim0.05$ would have Neptune trojans approach no closer than
$q\sim28.5$~AU (further than the distance at which we discovered  and
tracked the plutino L4m02).

We are not alarmed by the lack of such a detection, because the fraction
of TNOs which are Neptune trojans is very small.
To quantify this, we built a strawman trojan model and `observed' it
through the CFEPS survey simulator.
The model trojans had $a$ within 0.2~AU of 30.2~AU, 
$e$ uniform from 0 to 0.08,
with ascending nodes and mean longitudes uniformly distributed.
Libration amplitudes $L_{11}$ were chosen between 0--40$^{\circ}$,
with the relative number of objects having each libration amplitude
increasing linearly from 0 to 40$^{\circ}$.
Half of the trojans were set to be trailing ($<\!\!\phi_{11}\!\!> = 300^\circ$)
rather than leading ($<\!\!\phi_{11}\!\!> = 60^\circ$).
The resonant argument $\phi_{11}$ was chosen with sinusoidal time
weighting with amplitude $\pm L_{11}$ around $<\!\!\phi_{11}\!\!>$, 
with $\omega$ then calculated to fulfil the resonant 
condition.
Since the literature lacks the information needed to estimate the
inclination distribution, we chose a hot population with a similar inclination
distribution to the non-Kozai plutinos ($\sigma=15^{\circ}$).
The $H_g$-magnitude distribution was fixed with $\alpha=0.8$, as 
estimated by \citet{ShepTruj2010ApJ}.

We used the simulator to determine the Trojan population that would give 
3 or more CFEPS detections (on average); this provides the 95\% confidence
limit for Poisson statistics.
This limit is
\begin{equation}
N_{\mathrm{trojans}} (H_g < 9.16) < 300 \; \; [\; \mathrm{95\% \; confidence} \; ] \; ,
\end{equation}
when stated for the same $H_g$ value as the other resonances we
study.
\citet{ShepTruj2010Sci} estimate that there are $\sim$400 Neptune 
trojans with radii$>$40~km;  
assuming a 5\% albedo, this corresponds to $H_g\sim$9.6.  
Scaling our population upper limit using $\alpha$=0.8 makes the CFEPS
upper limit $<$600 trojans with $D > 80$~km (95\% confidence), 
indicating that the non-detection of a Neptune trojan in CFEPS is not 
statistically alarming given the 400-Trojan estimate of 
\citet{ShepTruj2010Sci}.


\section{CFEPS comparison to a cosmogonic model} 

The CFEPS project has produced three data products, all of which can
be accessed at {\it http://www.cfeps.net}.
First, there is database of TNO photometry and astrometry for TNOs
(characterized and non-characterized) seen in the survey.
The characterized list is intimately linked to the second data
product: the Survey Simulator, described below.
Thirdly, one can obtain an orbital element distribution (called the L7
synthetic model) which is an empirically-determined orbital and $H$ 
distribution
which, when passed through our Survey Simulator, provides a distribution of 
detections statistically indistinguishable from the CFEPS
detections.

The true power of CFEPS is the ability to compare a proposed model 
(resulting from a cosmogonic simulation) to reality.
In order to decide how well a proposed Kuiper Belt orbital distribution 
matches the CFEPS data, one must not just compare the Kuiper Belt model to the 
L7 synthetic model.
This is because CFEPS (or any survey) will be biased toward or against
detections in particular parts of orbital parameter space; 
a model seemingly different from the the L7 synthetic model may 
be biased when `viewed' through the CFEPS pointing history and flux limits
into an acceptable match.
Similarly, models which appear to match some aspects of the L7 synthetic model 
may fail dramatically.
The only quantitative way to compare a model to the Kuiper Belt via the 
CFEPS survey is to pass the model through the L7 Survey 
Simulator and compare the distribution of simulated to real 
detections. 
As an example of this process, we here examine the results of a cosmogonic
simulation based on the Nice model of giant planet migration
\citep{Levetal2008}, in order to compare the simulated plutinos with 
the CFEPS plutino orbital distribution.
We chose this model because the plutino libration amplitudes were 
made available by the authors; providing such information is the 
state of the art in Kuiper Belt formation models and should become
the norm.

We begin with the plutino orbital elements
from the end of Run B of \citet{Levetal2008}, which are those emplaced
during the planet-migration process and then survive 1 Gyr
`erosion' process to eliminate TNOs that did not have long-term
stability on the time scale of the Solar System's age.
Because there are only 186 model surviving plutinos, we create new 
particles with very similar orbital elements by ``smearing out'' those 
of the existing particles; 
values of $a$, $e$, and $i$ for each new particle were randomly chosen
within $\pm$0.1~AU, 0.02, and 5$^{\circ}$ of the orbital elements of
one of the original Nice model particles.
We verified that this does not change the overall shape of the cumulative
distributions for these orbital elements.
Next, $\phi_{32}$ is chosen sinusoidally from within the values allowed by
the known libration amplitude of the Nice model particle.
The ascending node's longitude $\Omega$ and mean anomaly ${\cal M}$ are chosen 
randomly, leaving $\omega$ to be chosen
to satisfy the resonance condition.
Lastly, the particle's $H_g$ magnitude is chosen from the same $\alpha$=0.9 
exponential distribution used for CFEPS plutinos.
The CFEPS survey simulator then evaluates whether or not it was 
detected.

\begin{figure}
\centering
\includegraphics[scale=0.5]{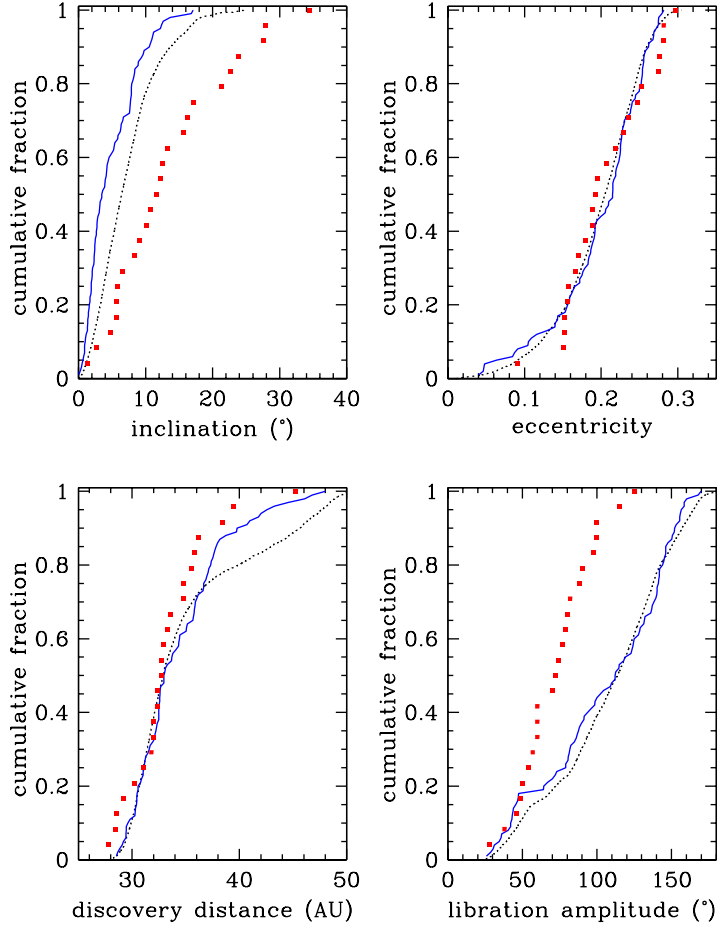}
\caption{
Comparison between CFEPS plutino detections and simulated detections from 
the Nice model plutino distribution. 
Red squares are real CFEPS plutino detections, the dotted black line shows the
intrinsic Nice model plutino distributions, and the blue line is the simulated
detections after running this intrinsic population through the CFEPS survey
simulator.
The magnitude distribution is not shown; this was not provided in the Nice
model data but we find using the same $H_g$ magnitude 
distribution as for the CFEPS plutinos produced an acceptable match (which 
is unsurprising given that the $e$ distribution is similar and $\alpha=0.9$
was chosen to represent the CFEPS detections).
While the eccentricity and discovery distance distributions match 
the CFEPS data reasonably well, 
the Anderson-Darling analysis indicates the CFEPS $i$ and $L_{32}$
distribution would occur $<$0.1\% of the time.
}
\label{fig:lev2008dist}
\end{figure}

The process was repeated until 10,000 synthetic detections were generated,
creating cumulative detection distributions  
(Fig.~\ref{fig:lev2008dist})
from which the probability of drawing the detected CFEPS sample
is judged.
The detected $e$'s and discovery distances provide statistically-acceptable 
matches to the CFEPS detections.
In contrast, the hypotheses that the $i$ or $L_{32}$ libration amplitude 
distributions of the CFEPS detections could be drawn from this 
Nice model simulation both fail at $>$99.9\% confidence.
The $i$-distribution of the detections that would come from
an intrinsic plutino distribution produced by the Nice model is far
too cold, and the $L_{32}$ distribution contains too
many large libration-amplitude objects.

Although this model is rejected, this style of model shows the forefront
of what models must now provide in Kuiper Belt science.
That is, a cosmogonic model should produce TNO orbital distributions
for the entire Kuiper Belt, including resonant libration amplitudes
and determination of Kozai resonance occupation.
Comparison with the current TNO distribution can only 
really be performed if the cosmogonic simulation (which often focuses
on events in early Solar System history) is dynamically eroded for
the $\sim$4~Gyr interval to bring it to the present day.
The fact that the \citet{Levetal2008} simulations were eroded for 1~Gyr
instead of 4~Gyr might result in small changes to the libration 
amplitude distribution of the survivors, but is unlikely to resolve
the major discrepancy given that \citet{NesRoig2000} and \cite{TiscMal2009} 
show that the distribution only changes appreciably with order of magnitude
increases of timescale.

We note a large number of non-resonant particles surrounding the 3:2 (and
some other resonances) with low $e$ at the end of the Nice model 
simulations.
We presume these TNOs to be generated during the phase where
the neptunian eccentricity is shrinking rapidly, which causes the resonance
to narrow and `drop out' formerly-resonant particles on either side
of the resonance.
We call these the `beards' of the resonance in this model.
This features should be preserved in the Kuiper Belt if the resonances
had abundant low-$e$ particles in the resonances when Neptune's $e$
dropped, but these beards are not obviously present in the real Kuiper-Belt
distribution.  
We doubt this is a selection effect, but are unable to present a 
quantitative analysis with the current CFEPS sample size.

\section{Resonant Populations} 
\label{sec:pops}

This work provides for the first time absolute population estimates
for a large variety of trans-neptunian resonances, allowing population
comparisons to quantitatively de-biased data that takes into account
the myriad of observational selection effects.
While the ratio of various resonance populations have been identified
as potentially diagnostic -- for example, \citet{JLT96} already mention
using the 2:1/3:2 population ratio to constrain Neptune migrations via
models like \citet{Malhot1995}-- the debiasing of the selection
effects for the two resonances has never been done to the level of
detail presented here.
\citet{ChiangJordan2002} and \citet{ChiangDES2003} showed models producing
population ratios of resonances to each other (for example, the 5:2 to 
2:1) or of sub-islands inside the 2:1 to each other, but again lacked 
the ability to compare to a survey for which the longitude coverage
could be quantitatively de-biased for selection effects.
\citet{HahnMalhot2005} produced ratios between resonances and to the
main belt from a model in the context of an outward Neptune migration into a 
pre-existing `warm' ($e=0.1$) belt, while \citet{Levetal2008} produced a
model in which the Kuiper Belt was moved out to its current
location; both of these models were forced to make comparisons to surveys
that could account for biases in, at best, an approximate way.

\begin{figure}
\centering
\includegraphics[scale=0.55]{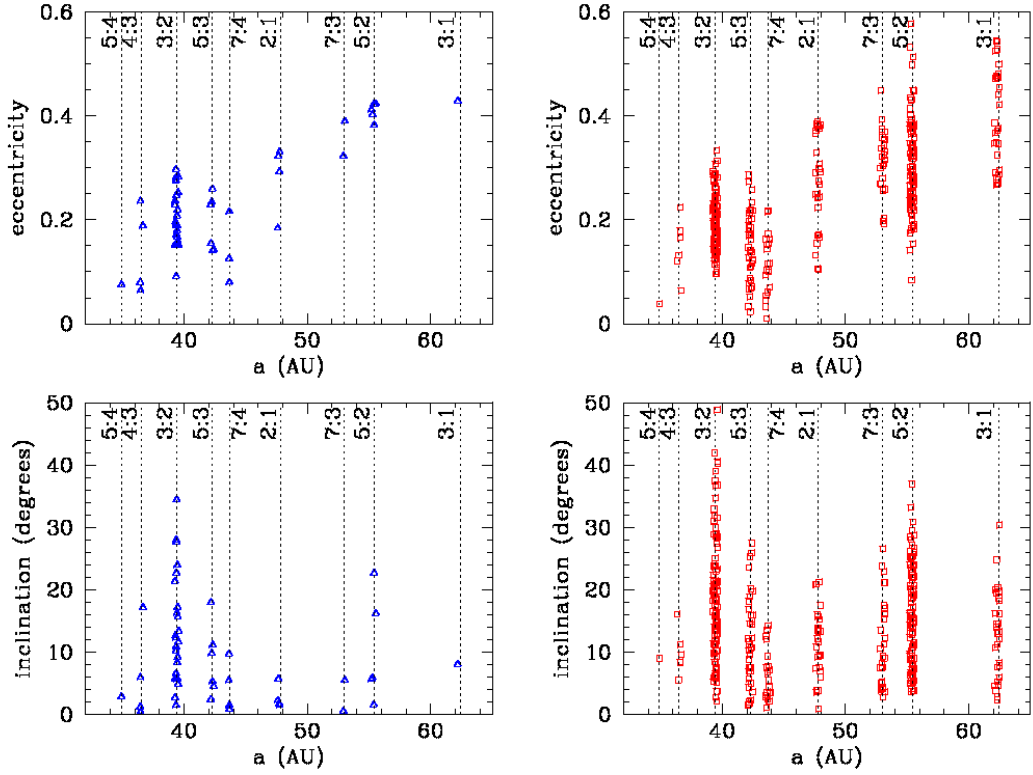}
\caption{The apparent versus debiased resonant Kuiper Belt.
The two left panels show the $(a,e)$ and $(a,i)$ distribution of 
the flux-limited CFEPS resonant detections from $a$=30--65~AU.
The right panels show the distribution of their debiased population,
scaled so that the plutinos have 100 members.
It is obvious that the true Kuiper belt has a higher fraction of 
larger-$a$, lower-$e$, and larger-$i$ members than the currently-detected 
sample.
The absence of low-$e$ resonant TNOs with $a>$46 AU is not absolutely
required by our modeling due to the detection biases against them.
}
\label{fig:aeipops}
\end{figure}

Fig.~\ref{fig:aeipops} shows the debiasing of the CFEPS, transforming 
the resonant populations from their biased apparent fractions (left
column) to their `true' values (right column, a debiased  sample from the
models presented in Table~\ref{tab:pop}).
An evident result is that the distant resonances make up a much
larger fraction of the total resonant population in reality than in the
flux-biased sample.
Although it is obvious the fraction of large-$a$ resonant TNOs (compared
to low-$a$ ones) will be higher in reality than in the flux-biased 
sample, this effect has never been quantified.
In particular it is obvious that beyond the
2:1 current surveys have just seen the `tip of the iceberg' and the
resonant populations contain many more large-$i$ and/or low-$e$
members than either CFEPS or the full MPC sample have yet exposed.

\begin{figure}
\centering
\includegraphics[scale=0.35]{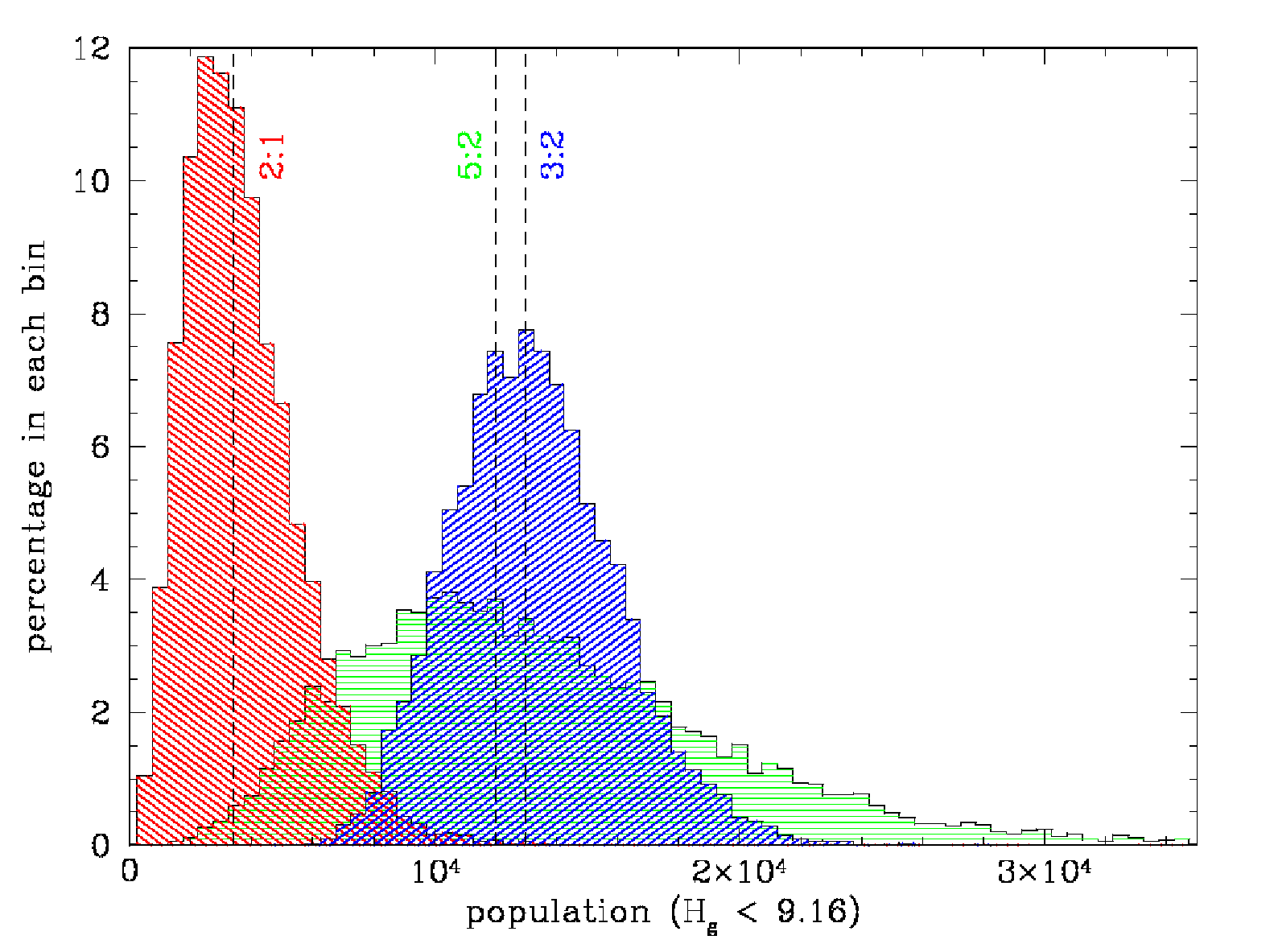}
\caption{Population estimates for the 3:2, 5:2, and 2:1 resonances.
Each histogram is a separately-normalized distribution of population
estimates which yield the correct number of detections for that 
resonance.  
Vertical lines show the median of the population estimates for each 
resonance.
Although the 2:1 and 5:2 histograms overlap, the probability that
the 2:1 population is larger than the 5:2 population when both are
randomly drawn from these distributions is $<$5\%.
}
\label{fig:pops}
\end{figure}

Population comparisons benefit from uncertainty estimates.
In particular, the population ratios in the well-studied 3:2, 2:1, and
5:2 are desirable.
To obtain a set of absolute population estimates, we drew particles at random
from our model orbital and $H$-magnitude distributions until obtaining
the true number of CFEPS detections for a given resonance
(Fig.~\ref{fig:pops}).
There is an essentially Poisson distribution of plausible
`true' populations that will allow the observed number of detections,
explaining the shape to the histograms in Fig.~\ref{fig:pops};
the median is reported in Table~\ref{tab:pop} along with the upper and
lower limits which leave only 2.5\% of the measurements in each 
tail.
Although we use conservative 95\% confidence regions (resulting in large 
stated uncertainties), CFEPS is for the first time able to provide measurements 
of the resonant populations that take into account the longitude
coverage and relative depth of its survey patches.

One of the most striking results (Fig.~\ref{fig:pops})
is that the best-estimate populations for the important resonance 
trio 3:2/2:1/5:2  are in the ratio $\sim$4/1/4.
To our knowledge, this is in stark contrast with all previously-published
models; those which obtain a weakly-populated 2:1 (relative to the 3:2)
never simultaneously have a 5:2 population equal to that of the
plutinos.
The simulations of \citet{ChiangDES2003} showed a huge 2:1/5:2 ratio
unless migration occurred into a hot disk which dropped the 
ratio to roughly 3/2 (to be compared to the 1/4 ratio we favor), with
a plutino population even larger than the 2:1. 
These authors ruled out `resonance sticking' of scattering TNOs as the 
dominant production method for 5:2 resonators due to the incorrect libration
amplitude distribution (a conclusion we share based on the small 
$L_{52}$ amplitudes for the CFEPS detections).
\citet{HahnMalhot2005}'s simulations into a warm primordial disk
exhibit 3:2/2:1/5:2 ratios of about 2/5/1, and
\citet{Levetal2008} produce 8/3/1.
That is, all simulations to date produce fewer 5:2 resonators than
twotinos by factors of several, whereas CFEPS indicates that the 
reverse is true (and rules out the 2:1 being more
populated than the 5:2 at $>$95\% confidence).
This is thus an important new constraint on formation models.

This kind of constant also holds for more distant resonances.
Both the 3:1 and 5:1 have best-estimate populations larger than
the 2:1 (although uncertainties are large), indicating that
large-$a$ resonant orbits must be efficiently populated by
formation models.
In models like \citet{Levetal2008}, where the Kuiper belt is
transplanted out, this is difficult to do because large-$a$ orbits
are inefficiently generated. 
The hypothesis that these objects are instead
swept into $a>50$~AU resonances from warm or hot populations located
at these distances before planet migration is faced with the problem of
explaining where all the non-resonant TNOs have gone.

This realization that the distant resonances are heavily
populated opens the possibility that the current scattering population
is dominantly being supplied by abundant resonant escapees.
If true then the resonant reservoir would be 
the ultimate source of Jupiter-family comets (JFCs), through 
the chain: resonant $\Rightarrow$ scattering $\Rightarrow$ Centaur
$\Rightarrow$ JFC.
Due to their chaotic boundaries,
the resonances provide a `leakier' source of scattering TNOs than the 
classical belt and most TNOs escaping a resonance would immediately 
find themselves on Neptune-coupled orbits and begin scattering\footnote{
\citet{Horner2010trojansmakejfcs} suggested that the Neptune Trojans
alone could be an important Centaur source, but it seems unlikely
that the other (vastly more populated) resonances would not dominate
the leakage supply.
}.
In such a scenario the escape rate from all resonances would 
balance the loss of actively scattering objects to the Centaur population or 
ejection from the Solar System.
The flaw in this scenario is that there seem to be
too many scattering TNOs in the current epoch to permit them
being anything other than the decaying remnant of a huge primordial
population \citep{dl1997}.
\citet{L7paper} estimate (to order of magnitude) that there are 
currently $\sim5,000$ $H_g<9.16$ actively-scattering TNOs 
(with a clarification on the definition of this population);
this is too large a fraction of the sum of the resonant populations
in Table~\ref{tab:pop} to permit the scattering population to be
in steady state.
\citet{VolkMalhot2008}
call into question even the `decaying remnant' scenario as the
supply rate they estimate from the metastable Kuiper belt (mostly a mix
of detached and resonant objects with $a>50$~AU and $q>$33~AU) 
into the Jupiter-Family comets seems too low given their extrapolation
of observational estimates of the `excited' ($i>$5$^\circ$) 
population in the 10--100~km size range. 
This analysis should be re-done however because Volk and Malhotra 
assumed that essentially all of today's `excited' TNOs (observed 
by various surveys) are scattered objects contributing to the 
Centaur supply chain, while 
in fact
the $a>50$~AU population has a very non-negligible resonant
component \citep{GladmanNomen2008}, and
\citet{L7paper} and this current manuscript show that the actively
scattering population is only a tiny fraction of the other
`excited' (resonant + hot classical + detached) populations.

The total resonant population is, however, also comparable to the
\citet{L7paper}
estimate for the sum of the outer classical and detached 
populations (of $\sim$80,000 with $H_g<9.16$).
This permits serious consideration of the hypothesis that most
detached TNO population are resonant objects that were
dropped out of resonance while the resonant objects were being emplaced,
but must generate roughly equal numbers of resonant and non-resonant objects 
surviving to the present day.

\section{Discussion}

Given the available constraints from the structure and relative
populations of various Kuiper Belt components, what can one conclude
about the processes that emplaced these components?
Based on our debiased understanding from CFEPS, we feel that the 
following constraints are of chief importance:

1. The resonant populations appear to be consistent with all being
   emplaced from a source population that lacked a cold component.
   (The differences between them can be plausibly explained by capture 
    or subsequent erosion processes that are inclination-dependent.) 

2. The inner classical belt and outer classical belt lack a 
   cold inclination component \citep{L7paper}, with only the 
   main belt having both hot and cold components.

3. The sum of the resonant populations is $\approx$75\%
   that of the main belt, for $H_g<8$.
   
4. The current `actively scattering' disk is $\sim$5\% of the main belt
   population, with at least factor of two uncertainty.

Although we do not support it here with detailed simulations, we believe
that the following scenario could explain the known structure.

A crucial feature is that the cold population is confined to the $a$=42-47~AU
region of the main belt, with a hotter $e$ distribution for $a>44.4$~AU.
We postulate the cold population could be primordial, with an initial
outer edge at this 44.4~AU boundary.
The plausible scenario consists of all the other Kuiper Belt populations
(hot classical, including the inner and outer belts, detached objects,
resonant objects, and the currently scattering objects) being planted into
the belt via a mechanism similar to that described by 
\citet{Gomes2003} and 
\citet{Levetal2008}, in which a massive scattering disk is flung out by
the migrating giant planets; resonant trapping of the scattering 
objects and subsequent dropout litters the hot classical population behind
the slowly-advancing resonances (which are wide and powerful due to 
Neptune's temporarily larger $e_N$).
Neptune `jumps' out several AU due to encounters with Uranus and both
planets decouple due to damping of their eccentricities. 
Today's resonant objects are those which were still trapped during the final
stages of this process as $e_N\rightarrow 0$.
Unlike the \citet{Levetal2008} model, we posit: 
(1) The scattering disk extends to very large $a$ already when Neptune 
    `jumped' out to nearly 30~AU.
(2) This scattering disk was very hot; essentially the  
    $\sigma_h\simeq15^\circ$
    width which all the non-cold populations share.   How this happens
    is unknown.
(3) The cold population is already in place; it is largely unaffected
    because the 2:1 resonance jumps to or beyond the 44.4 AU edge.  
    Eccentric Neptune is able to dimly `stir' the $a/e$ distribution of
    the cold disk, keeping most stirred perihelia $q<44$~AU, before $e_N$
    rapidly decays.

A critical constraint is to prevent TNOs from the cold population appearing
in either the 3:2 or 2:1 resonance; this requires that after jumping to
near $a$=30~AU, any remaining small outward Neptune migration cannot allow 
the resonances to sweep through a cold population, because it would readily
trap and preserve them \citep{HahnMalhot2005}.
Keeping the 2:1 free of low-$i$ TNOs can be accomplished by having the
post-jump value of the resonant semimajor axis beyond the outer edge of
the cold disk (say, landing in the 45--46~AU range before finishing
outward migration by another ~AU or so).
The situation with the 3:2 is more complex because its lack of a cold
component seemingly implies that by the time Neptune jumped, the 
semimajor axis range  between the post-jump $a_{3:2}\sim37$ AU and today's 
value must have already been empty of cold objects.  
Although a primordial inner edge of the cold population is not impossible,
the fact that the current $a=42.4$~AU inner boundary of the cold population
is at the border of the $\nu_8$ secular resonance allows a scenario 
in which this strongly-unstable secular resonance swept through the 
$\sim$37-39~AU region prior to any final small-distance Neptune migration; 
\citet{holwisdom1993} show
that the $\nu_8$ drives particles to Neptune encounters in only
$\sim30$~Myr, which is comparable to the migration time scale for
Neptune in \citet{Levetal2008}.
In this scenario, the primordial cold objects with $a<42.4$~AU join the 
scattering TNOs, but make up only a tiny fraction of this population
as they are `diluted' if any of them are later re-planted into the
Kuiper Belt.
Unfortunately, the timing (and even migration direction) of the
$\nu_8$ is unclear; 
\citet{NagIda2000} show early and rapid migration of the $\nu_8$
inwards as the the protoplanet disk's mass eroded, but their
calculations did not include the probable outward migration of 
Neptune.

In our scenario one has an easy explanation for the differences in
colours, size distribution, and binary fraction of the cold main-belt
fraction; the cold belt was simply steeper, redder, and either
formed more binaries or preserved a greater fraction of them, unlike
the implanted components \citep{ParkerKav2010}.
Although there is no direct observational timing constraint, this implantation
scenario seems more natural if the disk is scattered very early in
the Solar System's history, without the $\sim$600~Myr delay proposed
in the  Nice Model \citep{GomesEtal2005}.
In fact, our scenario does not stipulate where the `early' scattering 
component comes from, although the most plausible source is 
it being perturbed out from the planetesimal-rich giant-planet region 
interior to 30~AU.  
At the time of Neptune's jump, this early scattering population
must extend to $a>$50~AU in order to allow efficient trapping into 
the well-stocked distant resonances.

The mechanism that causes this early scattering population (which is the source
for all hot Kuiper Belt populations) to have the needed inclination width
of $\sigma\sim15^\circ$ is unclear. 
Perhaps the giant planets somehow vertically heated the planetesimal belt 
before it was scattered out (although in general scattering will pump
$a$ and $e$ at least as fast as $i$).
\citet{Gomes2003} manages to produce large-$i$ implantations from a
source disk, although the more recent \citet{Levetal2008} study 
produced much colder implanted population.
Perhaps other now-gone (`rogue') planets caused the initial vertical 
dispersion, although this too seems inefficient \citep{gladchan2006}.
Very nearby stellar encounters could generate the inclinations
by scattering objects
\citep[{\it eg.}][]{Kobayashi2005}
but preserving the $\sigma\sim2^\circ$ cold disk in a $\sim$44~AU 
ring is a very strong constraint.

The following estimates of sub-populations are intended only to provide
a coherent picture to a factor of 3 or so, with
all population estimates for $H_g<9.16$ (roughly $D>$100~km).
CFEPS estimates \citep{L7paper} that today's scattering population 
is $\sim10^4$. 
Assuming this is {\it not} currently in steady state re-supply from 
another source, \citet{dl1997} estimate that this would require 
about $\sim$100 times as many scattering objects $\sim$4~Gyr ago; 
in a scenario where this disk goes to considerably smaller perihelion
distances than the current $q\sim35$~AU, the initial population would
have been at least several times larger and we take $10^7$ initial
$D>$100-km scattering bodies.  
In an $\alpha\simeq$0.8 size distribution most of the mass in in the
small end, and the resulting $\sim$10~$M_\oplus$ of bodies is comfortably
smaller than the mass of the outer planets.
We take this primordial scattering population to be the source of the
high-$i$ populations.
\citet{Levetal2008} estimate $\sim$0.5\% of such a primordial scattering 
gets trapped into non-resonant orbits, implying a hot classical population
of $\sim50,000$, which is comparable to the 35,000 estimated in
\citet{L7paper} when one realizes that it is only the hot main-belt
population that is relevant (the cold population being pre-existing
in our scenario).
In addition, \citet{Levetal2008} report that the plutinos make up about
20\% of the non-resonant objects implanted in the main belt, or about
10,000 objects, again reasonably in accordance with the CFEPS estimate
of 13,000.
This scenario is not here supported by simulations, which would need 
to show that 
(1) the cold belt could survive the process, 
(2) the distant resonances can be efficiently filled, and 
(3) the \citet{Levetal2008} trapping fractions are not strongly affected 
by the hotter primordial scattering population that is required.
In this scenario, gradual migration is a relatively unimportant 
process for the Kuiper Belt's current structure.

Much of the excitement in Kuiper Belt studies comes from the 
vigorous interplay over the last two decades between observation
and theory, and the steady stream of unexpected discoveries
in both domains.
Much work remains to be done.
While there is evidently considerable room for future surveys to improve 
upon the CFEPS estimates, this can only be done with well-characterized
surveys whose selection effects are rigorously monitored.
In turn, the debaised orbital elements distributions will lead to 
much tighter constraints on models seeking to solve puzzles still present in 
our understanding of how the outer Solar System settled to its current state.





\begin{deluxetable}{c|cccc|cc} 																									
\tablecaption{Resonance Populations.																									
\label{tab:pop}}																									
\tablehead{																									
Res.	&	\# of	&	e$_c$	&	e$_w$	&	$\sigma_i$			&	Median Pop.						&	Median Pop.						\\
	&	det.	&		&		&	($^\circ$)			&	($H_g < 9.16$)						&	($H_g < 8$)						} 
\startdata																									
3:2	&	24	&	0.18$_k$	&	0.06$_k$	&	16$_k$	$^{+8}	_{-4}$	&	13,000	$^{+	6,000	}_{-	5,000	}$	&	1,200	$^{+	500	}_{-	400	}$	\\
5:2	&	5	&	0.30	&	0.10	&	14	$^{+20}	_{-7}$	&	12,000	$^{+	15,000	}_{-	8,000	}$	&	1,100	$^{+	,1400	}_{-	700	}$	\\ \hline
4:3	&	4	&	0.12	&	0.06	&	8	$^{+6}	_{-3}$	&	800	$^{+	1,100	}_{-	600	}$	&	70	$^{+	100	}_{-	50	}$	\\
5:3	&	6	&	0.16	&	0.06	&	11	$^{+14}	_{-5}$	&	5,000	$^{+	5,200	}_{-	3,000	}$	&	450	$^{+	470	}_{-	280	}$	\\
7:3	&	2	&	0.30	&	0.06	&	$\sim$10			&	4,000	$^{+	8,000	}_{-	3,000	}$	&	320	$^{+	760	}_{-	270	}$	\\ \hline
5:4	&	1	&	0.12	&	0.06	&	$\sim$10			&	160	$^{+	700	}_{-	140	}$	&	10	$^{+	60	}_{-	9	}$	\\
7:4	&	5	&	0.12	&	0.06	&	5	$^{+9}	_{-3}$	&	3,000	$^{+	4,000	}_{-	2,000	}$	&	300	$^{+	400	}_{-	200	}$	\\ \hline
2:1	&	5	&	0.1-0.4	&	-	&	7	$^{+0.5}	_{-5.5}$	&	3,700	$^{+	4,400	}_{-	2,400	}$	&	340	$^{+	400	}_{-	220	}$	\\
3:1	&	1	&	0.25-0.55	&	-	&	$\sim$10			&	4,000	$^{+	9,000	}_{-	3,000	}$	&	340	$^{+	800	}_{-	290	}$	\\
5:1	&	1	&	0.35-0.65	&	-	&	$\sim$10			&	8,000	$^{+	34,000	}_{-	7,000	}$	&	700	$^{+	3,000	}_{-	700	}$	\\

\enddata
\tablecomments{
Principle parameters for the models of each mean-motion resonance.
All resonances used $\alpha=0.9$ for the $H_g$-magnitude distribution (that
measured for the plutinos).
Uncertainties reflect 95\% confidence ranges.
Population estimates for $H_g<9.16$ correspond to 100-km diameter (for
nominal albedo), while $H_g<8$ estimates are provided for comparison
with the classical-belt population estimates of
\citet{L7paper}.
The $k$ subscript for the plutinos indicates that these are the parameters
for the non-Kozai component.}
\end{deluxetable}

\newpage

\section{Appendix}

In order to measure the CFEPS bias to get an estimate of a resonance's
true population, we select model objects by randomly drawing from a 
parametrization of the orbital distribution for the given resonance,
and assigning an $H_g$ magnitude from a power-law distribution.
Each object is then run through the CFEPS survey simulator to decide whether
or not it was detectable.  
This is repeated until a requested number of detections is reached;
this number is usually either  (a) $\sim10^4$ to obtain a well-sampled
distribution of orbits that would be detected if the model was correct or
(b) the number of CFEPS detections to get an estimate of the true absolute
population of that resonance.
In case (a) the orbital distribution of the simulated detections is then
statistically compared to that of the real detections to decide whether 
or not that model is reasonable.

The orbital elements for each object are chosen in a different order 
depending on which resonance the object is a member of.  
This is because of the differing internal constraints of each resonance.  
The plutinos have many detections, allowing a much more in-depth exploration
of the possible orbital parameter distributions, as well as having a 
significant Kozai fraction (Section \ref{sec:plutsim}).  
The n:1 resonances have symmetric and asymmetric libration islands which 
must be populated (Section \ref{sec:n1sim}).
Other remaining resonances have fewer detections, and thus the model
need not be as complex and the orbital element distribution cannot be 
constrained as well.
These selection processes is described below, in order of increasing
complexity.

\subsection{Simulating the 5:2, n:3, and n:4 populations} \label{sec:otherorb}

Each of these resonances has between one and six CFEPS detections 
(Tables~\ref{tab:elems} and \ref{tab:elems2}), 
allowing population estimates but no detailed modelling of orbital
element distributions.  
The orbital elements and magnitudes of the synthetic objects in 
each of these resonances are chosen in the following order:

First the eccentricity is chosen  randomly from a Gaussian distribution 
centered on the input parameter $e_c$ with a width $e_w$.
Negative eccentricities and those that cause the object to approach the 
orbit of Uranus (q $\sim$ 22 AU) are  redrawn.
The semi-major axis is then chosen.  
This is drawn randomly within 0.2 AU of the resonance center.
Although in reality the resonances have semimajor axes boundaries that are
$e$ dependent, the effect on observability is so  weak that given
the numbers of detections and the fact that the $e$ distribution is
strongly peaked, this makes no difference to our current estimates.

The inclination is chosen independently of $a$ and $e$.
We use the $i$-distribution parametrization where the probability
of a given $i$ is $\propto \sin i\;e^{-i^2/2\sigma^2}$ as proposed by
\citet{Brown2001}.
The ascending node $\Omega$ and mean anomaly ${\cal M}$ are chosen 
randomly from 0--360$^\circ$.  

The libration amplitude $L$ for each TNO is chosen from a tent-shaped
distribution based on the plutino libration amplitude distribution suggested 
by \citet{LykMuk2007}.  
However our study of the plutinos leads us to use a slightly asymmetric
shape (see Sec.~\ref{sec:plutLibAmps}).
Our smallest libration-amplitude TNOs have $L=20^\circ$, the largest 
have $L=130^\circ$, and we put a peak in the libration amplitude distribution 
at 95$^\circ$ (that is,
the probability increases linearly from 20--95$^\circ$ and then
drops linearly to zero probability at 
$L=130^\circ$).

After a libration amplitude is chosen, $\phi_{jk}$ is chosen sinusoidally
within the range allowed by the libration amplitude (that is,  $\phi_{jk}$  
= $L\sin\psi$ where $\psi$ is a random phase).
The argument of pericenter $\omega$ is calculated via
$\phi_{jk}$~=~$j\lambda$~-~$k\lambda_N$~-~($j$-$k$)~$\varpi$.
Finally, the $H_g$ magnitude is chosen from a power-law distribution
10$^{\alpha H}$ with a maximum $H_g$ of 11; because this is well below the 
CFEPS detection limit, our estimate has no dependence on the cutoff.

As each object is generated, its orbital elements and $H_g$ are passed to 
the CFEPS Survey Simulator, which evaluates its detectability.
If it falls within one of the CFEPS pointings and is bright enough, it becomes
a synthetic detection.
These detections include a certain fraction of objects that will 
be ``lost'' due to tracking losses in a magnitude dependent way 
\citep[see][]{L7paper}.

After the desired number of synthetic detections have been acquired,
the distribution of synthetic detections and real detections are compared
in five parameters: inclination, eccentricity, distance at detection,
apparent g magnitude, and libration amplitude (see Figure~\ref{fig:cumuplots32}),
as discussed in \citet{L3paper}.
The Anderson-Darling statistic 
of the CFEPS detections
relative to the simulated detections are calculated for each distributions
and the probability of a departure as large or larger than  the detected
sample is determined by bootstrapping each sample.  
We consider a model rejectable when at least one of the five distributions
has a bootstrapped probability of $<0.05$.

For these resonances there are insufficient detections to constrain
the orbital distribution directly, but this does not result in an
important uncertainty in the population estimate.
For example, modelling the 5:2 libration amplitude distribution
as flat from 0--130$^\circ$ does not result in a rejectable model,
like it did for the plutinos, but the $H_g<9.16$ population 
only drops to 11,000 from 12,000 (Table~\ref{tab:pop}), a change
vastly smaller than the uncertainties) thus showing 
that our 5:2 population estimate is insensitive to the unknown 
libration-amplitude distribution.
As a second example, changing the $e_w$ value for the 4:3 resonance
from 0.06 to 0.10 (allowing easier-to-detect higher-$e$ resonators
to exist) drops the $H_g<9.16$ best estimate from 800 (+1100,-600)
to 640.
We thus believe our Poisson uncertainties due to small numbers
of detections dwarf the systematic errors for resonances other
than the 3:2.

\subsection{Simulating the Plutino population} \label{sec:plutsim}

The model for TNOs in the 3:2 resonance is identical to the previous
section, except that for this resonance we also force a fraction $f_K$
of the objects to simultaneously be in the Kozai resonance.
The presence of the Kozai resonance inside the 3:2 is well studied
\citep{MorbThomMoons, NesRoig2000, WanHuang2007}.
While the Kozai resonance appears only at very large inclinations and
eccentricities for TNOs outside mean-motion resonances \citep{ThomMorbKoz},
inside the 3:2 mean-motion resonance the precession rates rise enough
that at moderate ($e\sim0.25$) and inclination ($i\simeq$10--25$^\circ$)
the Kozai effect causes libration of the TNO's argument of perihelion
around $\omega=90^\circ$ or $270^\circ$,
which results in its perihelion direction being barred from the
plane of the Solar System.

\begin{figure}
\centering
\includegraphics[scale=0.5]{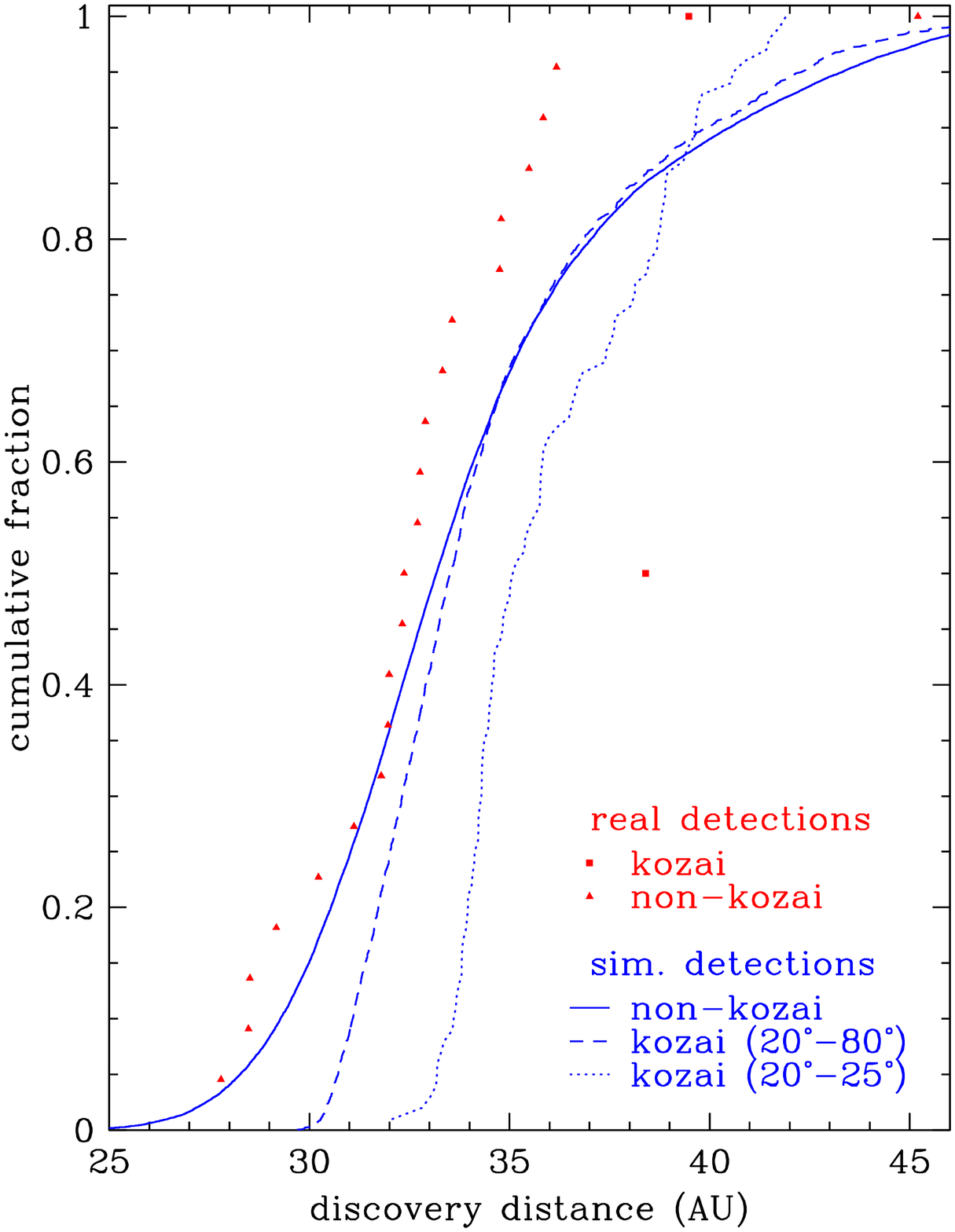}
\caption{Model predictions for the detection distance distribution for non-Kozai plutinos,
Kozai plutinos with $\omega$ libration amplitudes in the range 20--80$^{\circ}$,
and Kozai plutinos with libration amplitudes restricted to the range
20--25$^{\circ}$.  
The cumulative distribution of the real CFEPS detections is also shown.
Kozai plutinos, especially those with small libration amplitudes, are preferentially 
detected at larger distances.
}
\label{fig:kozaidist}
\end{figure}

Two of the 24 CFEPS-detected plutinos are in the Kozai resonance, and 
the plutinos were already known to include a significant Kozai component 
\citep{LykMuk2007}.
The fraction of Kozai librators in the sample is one of our model
input parameters.  
One effect on the detection of plutinos in an ecliptic survey like CFEPS 
is that Kozai librators are preferentially detected at larger distances
than non-Kozais (Figure~\ref{fig:kozaidist}).

During model construction, each object is labeled as either a Kozai or 
non-Kozai resonator using the model's value of $f_K$,  
with the goal being that the simulated detected fraction is satisfactorily in
agreement with the true detection fraction of 2/24.
If the object is not in the Kozai resonance, the orbital
parameters are chosen as described in Sec.~\ref{sec:otherorb} above, 
with a slight change to the way the semi-major axes are chosen.
Instead of just choosing them randomly within 0.2 AU of the 
center of the resonance, 
following Fig.~7 of \citet{TiscMal2009}, we narrow the resonance's
$a$ width linearly to zero as $e$ drops from 0.16 to 0.01; if the
drawn ($a,e$) pair falls outside this bound a new $a$ and $e$
are drawn.

\begin{figure}
\centering
\includegraphics[scale=0.5]{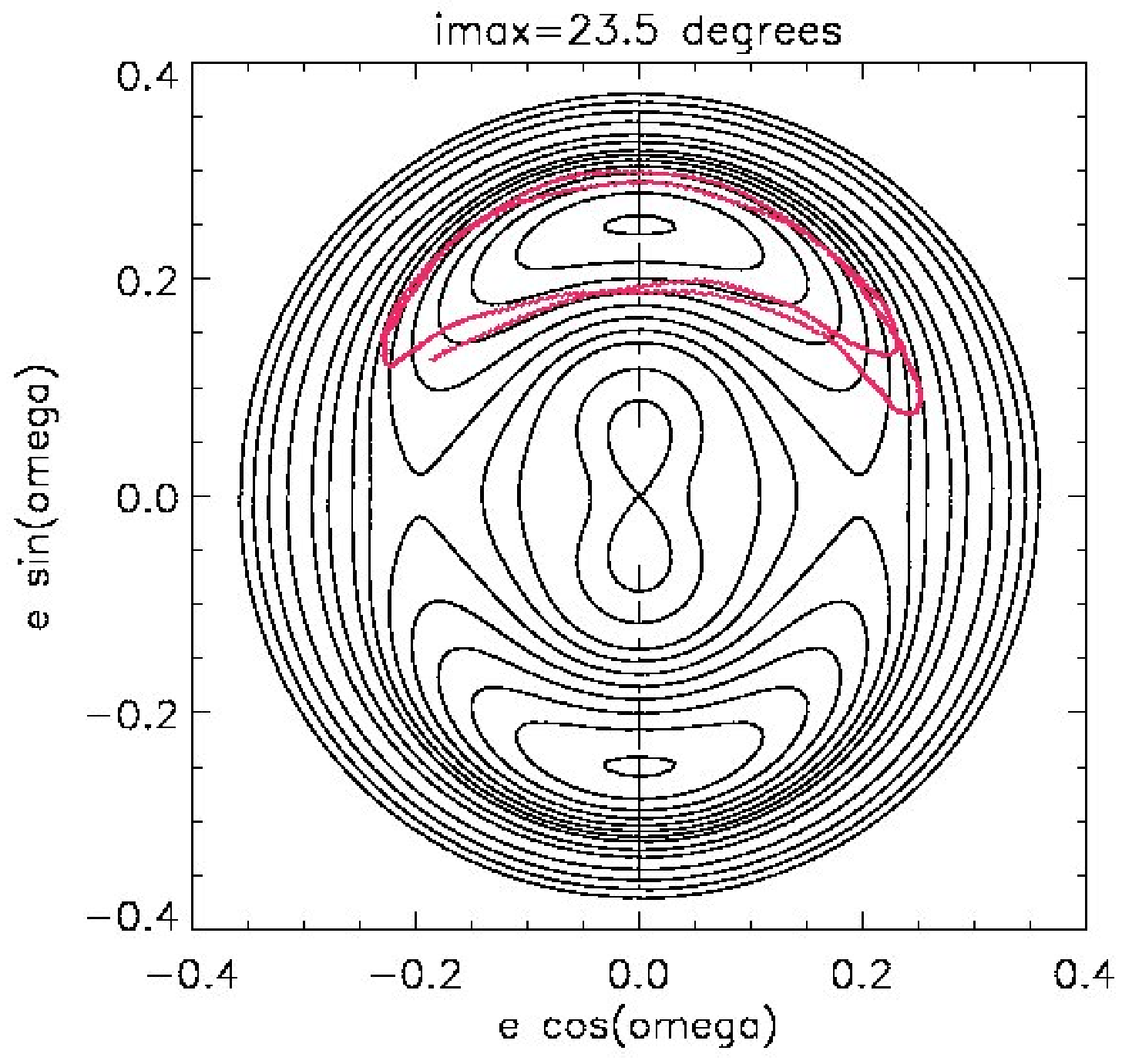}
\caption{The Hamiltonian phase space for the set of Kozai
librators used in the CFEPS-L7 plutino model.
Here eccentricity $e$ is the radial coordinate and the polar 
angle is $\omega$.
This diagram's set of contours corresponds to the angular momentum where
the zero-eccentricity orbit has $i_{max}=23.5^\circ$.
Also shown (overlain) 
is the trajectory of a 10-Myr integration of a real Kozai plutino 
(numbered TNO 69986).
}
\label{fig:kozaiphase}
\end{figure}

For the fraction $f_K$ of the plutinos chosen to be Kozai resonators, 
the following procedure for choosing orbital elements is followed:

First, a Hamiltonian level surface was generated, based on the 
calculations of \citet{WanHuang2007}.
The libration trajectories in $(e,\omega)$ space are determined by
the value of the $z$-component of the angular momentum, which is 
equivalently labelled by value of $\cos i_{max}$ for the circular
orbit of the same angular momentum.
For our current purposes, we picked the single value of 
$i_{max}$ = 23.5$^{\circ}$ based on visual
comparison with integrations of known plutino Kozai 
librators (Figure~\ref{fig:kozaiphase}).  
With this fixed, a libration trajectory is picked at random, corresponding to 
Kozai libration amplitudes between 20 and 80$^{\circ}$.  
$\omega$ is picked sinusoidally between 90$^{\circ}$ and the 
maximum $\omega$ allowed by the chosen contour.  
The eccentricity is then found numerically using $\omega$,
the chosen Hamiltonian trajectory, and equation (9) from 
\citet{WanHuang2007}.
Then $i$ is calculated using conservation of the z-component of angular 
momentum 
($L_z \propto \cos i \sqrt{1-e^2}$).
We then move half the Kozai librators to be around the 
$<\omega>=270^\circ$ island by the transformation
$\omega \longrightarrow 360^{\circ} - \omega$.  

Next, the semimajor axis is chosen in the same manner as for 
the non-Kozai plutinos, and ${\cal M}$ is chosen randomly.
Libration amplitudes for $\phi_{32}$ are chosen in the same way as for the 
non-Kozais, again using the tent-shaped distribution.
$\phi_{32}$ itself is chosen sinusoidally within the values allowed
by the chosen libration amplitude, which this allows 
the ascending node $\Omega$ to be calculated again using
the relation 
$\phi_{32}$ = $3\lambda$ - $2\lambda_N$ - $\varpi$.

All the orbital elements have at this point been chosen, so 
after choosing a $H_g$ magnitude from the same power-law 
distribution, the object is completely defined
and is sent to the survey simulator to evaluate 
whether or not is will be counted as a synthetic detection.

\subsection{Simulating the n:1 populations} \label{sec:n1sim}

Because the $n$:1 resonances possess several libration islands,
the intrinsic orbital distribution must be picked in a more
complex way before it is passed into the survey simulator.
Compared to the plutinos and 5:2, we have far fewer CFEPS objects in these 
resonances than are needed to directly constrain their complex
internal structure.
Thus, our primary goal is to obtain a calibrated absolute population
estimate based on the expected internal structure predicted by
analytic studies of these resonances.

\begin{figure}
\centering
\includegraphics[scale=0.4]{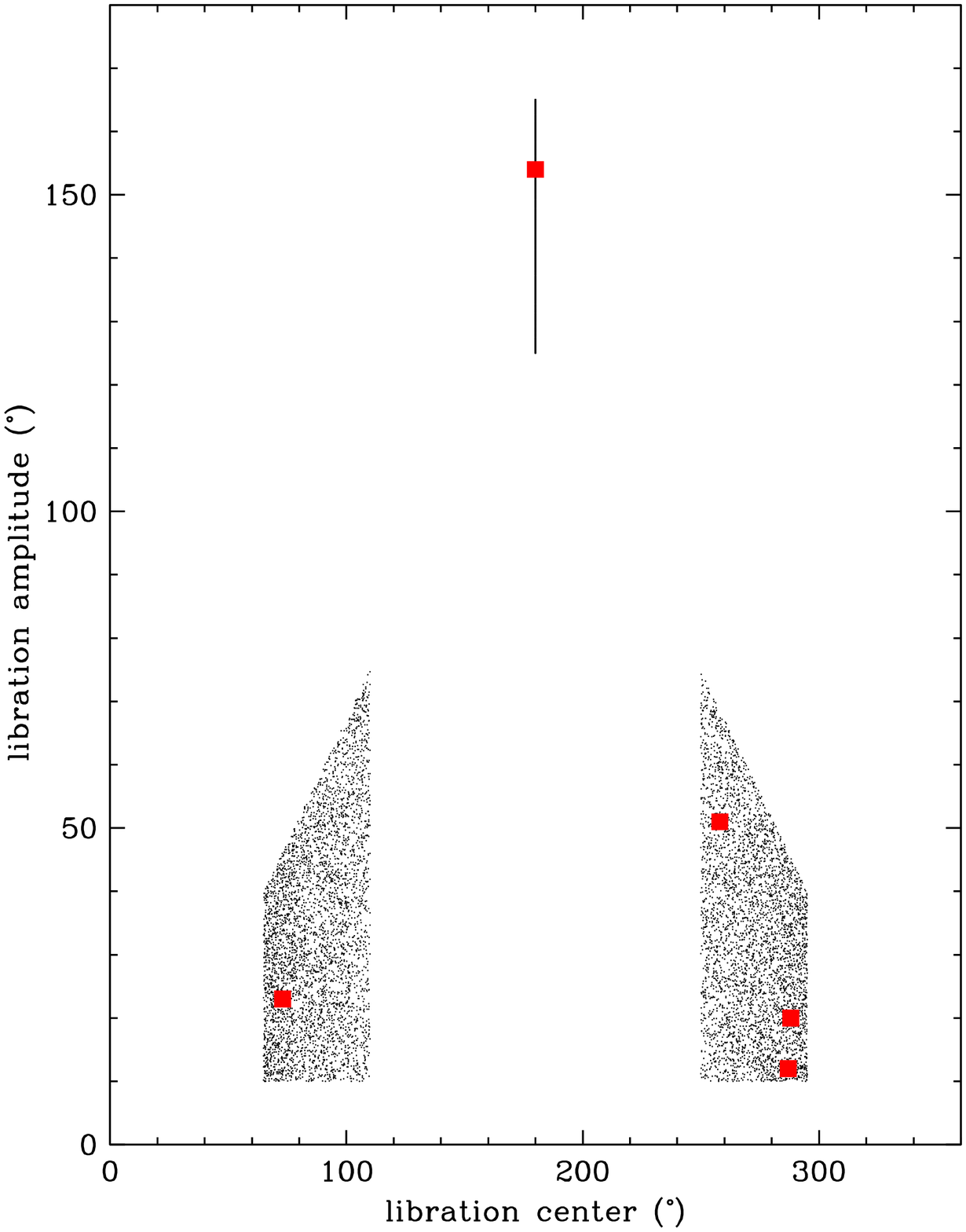}\includegraphics[scale=0.4]{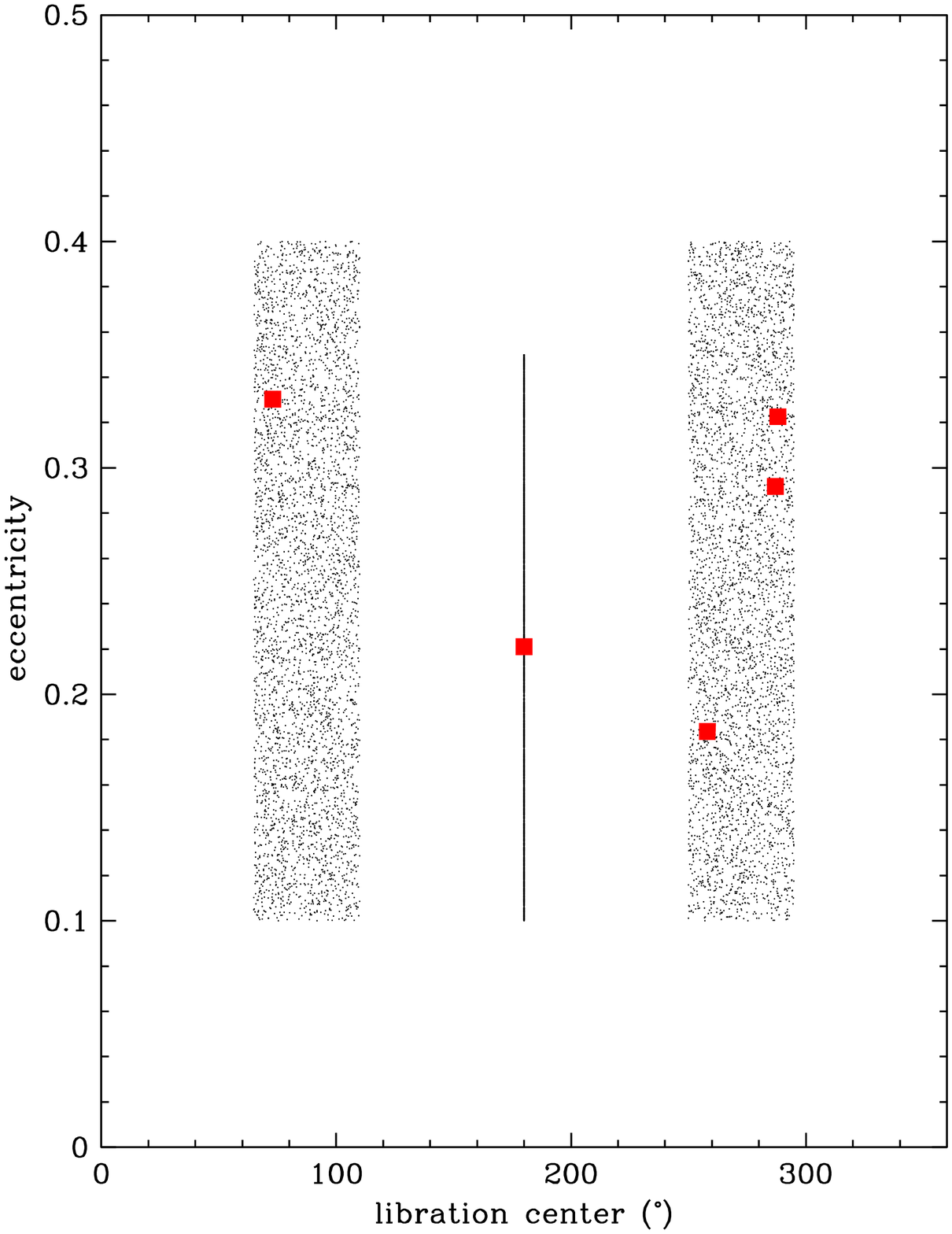}
\caption{
The range of libration amplitudes, libration centers, and
eccentricities chosen in our model for the symmetric and asymmetric islands
in the 2:1 resonances.  The 3:1 and 5:1 models are the same except
for a differing range in eccentricities (see text).
Real 2:1 detections are shown as red squares.
}
\label{fig:lib21}
\end{figure}

Because the effect on observability of the $a$-width of the resonance
is tiny, we pick the semi-major axis for model  $n$:1 resonant TNOs
randomly within 0.2 AU of the resonance center.
The eccentricity distribution is more complex because it is linked 
to the structure of the asymmetric islands.
We have incorporated the main features of the resonance from studies
of the structure and erosion (see, for example,
\citet{ChiangJordan2002} and \citet{TiscMal2009}).
We define the symmetric fraction $f_s$=30\% for each $n$:1 resonance to be 
the fraction which are librating in the symmetric island, and as a 
working hypothesis take the remaining objects to be evenly divided 
between the two asymmetric libration islands.
Symmetric librators have a resonant argument 
$\phi_{n1}$ = $n\lambda - \lambda_N - (n-1)\varpi $  which librates around
$<\!\!\phi_{n1}\!\!> = 180^\circ$ with amplitudes $L_{n1}$ ranging from 
125--165$^\circ$ (Fig.~\ref{fig:lib21}), while the asymmetric librators 
have a more complex distribution.
`Leading librators' 
(to use the terminology of Chiang and Jordan, denoting orbits
whose pericenter directions are somewhat ahead of Neptune)
are randomly given libration centers $<\!\!\phi_{n1}\!\!>$ in the interval
65--110$^\circ$,  with libration amplitudes $L_{n1}$ from 
10--75$^\circ$, where we redraw if $L_{n1}$ is greater than a 
limit which linearly rises from $L_{n1}$=40$^\circ$ for 
$<\!\!\phi_{n1}\!\!>=65^\circ$ 
to $L_{n1}=75^\circ$ for objects with 110$^\circ$ 
libration centers (Figure ~\ref{fig:lib21}). 
This range sufficiently reproduces the main characteristics of analytic 
studies of the asymmetric islands \citep{Beauge1994},
of numerical results on the post-migration distribution 
\citep{ChiangJordan2002},
and of the known 2:1 detections. 
Half of the asymmetric librators then have their centers moved to the
`trailing island' via 
$<\!\!\phi_{n1}\!\!> \longrightarrow 360^\circ \; - <\!\!\phi_{n1}\!\!>$.
Eccentricities for 2:1 resonators are drawn uniformly in the range 
0.10--0.35 for symmetric librators or 0.10--0.40 for 
asymmetric librators \citep{ChiangJordan2002}.
For the 3:1 the symmetric/asymmetric $e$ range is 0.25--0.50/0.25--0.55,
and for the 5:1 they are 0.35--0.60/0.35--0.65.
The dependence of the population estimates on the $e$ range chosen
is small; if the 2:1 eccentricity distribution is changed to be
uniform from 0--0.35 for all three islands, the model's rejectability
is not altered (Anderson-Darling $e$ match changes negligibly from 
0.69 to 0.47) and the population rises from 3700 (+4400,-2400) to
5700 due to the greater preponderance of harder-to-detect low-$e$
2:1 resonators in this alternate model.  
While this test is somewhat artificial because such low-$e$ twotinos are 
not abundantly present in \citet{ChiangJordan2002} or \citet{TiscMal2009},
even this large systematic change only alters the population estimate by 
half of our estimated uncertainty range.

Inclinations are chosen from a $\sin(i)$ times a gaussian
distribution (as for non-Kozai plutinos and for the other
resonances).
${\cal M}$ and $\Omega$ are chosen randomly from 0--360$^\circ$.
$\phi_{n1}$ is chosen sinusoidally from within the range of 
possible libration amplitudes around the libration center.
Then $\omega$ is calculated using the relation
$\phi_{n1}$ = $2{\cal M}$ + $\Omega$ + $\omega$ - $\lambda_N$.

Finally, the $n:1$ object is assigned an $H_g$ magnitude in the same manner 
as for the other resonances, regardless of which libration island it is 
assigned to.
It is then sent to the survey simulator.

\section{Acknowledgements}

We acknowledge the research support of the National Sciences and Engineering Research Council and the Canadian Foundation for Innovation.
We thank the queued service observing operations team at CFHT for their
excellence in obtaining the CFEPS observations.
J.~Hahn and H.~Levison provided output from their cosmogonic models,
thus allowing the quantitative comparisons between models and observation 
that are the future of Kuiper Belt studies.
Lastly, we acknowledge the historical debt this subject owes to Brian Marsden,
and honor his dedication to orbital computation.


\end{document}